\documentclass[final,3p,11pt]{elsarticle}

\usepackage[switch]{lineno} 
\usepackage[noend]{algpseudocode}
\usepackage{lipsum}
\usepackage{algorithm}
\usepackage{algorithmicx}
\usepackage{printlen}
\usepackage{etoolbox}

\usepackage{booktabs,multirow}
\usepackage[hidelinks]{hyperref}
\usepackage{upgreek}
\usepackage{comment}
\usepackage{optidef}
\usepackage{graphicx}   % For image handling
\usepackage{subcaption} % For subfigure environments
\usepackage{booktabs}   % in your preamble
\usepackage{bm}
\usepackage{media9}

\usepackage{xcolor}
\usepackage{soul}
\definecolor{myyellow}{HTML}{FFFFCC}
\definecolor{mygreen}{HTML}{CCFFCC}
\definecolor{mypink}{HTML}{FFE4E1}
\definecolor{mylightblue}{HTML}{B2FFFF}

\AtBeginEnvironment{algorithmic}{\lineskip0pt}

\usepackage{enumitem}
\setlist[enumerate,1]{label=\arabic*}
\setlist[enumerate,2]{label=\theenumi.\arabic*}
\setlist[enumerate,3]{label=\theenumii.\arabic*}
\journal{Ocean Engineering}
\usepackage{tikz}

\begin{document}
%\tikz[overlay] \node [fill=mylightblue,draw=mylightblue,thick,minimum width=6.3in,minimum height=0.49in] at (3in,-0.49in) {}; % title highlight

\begin{frontmatter}

%\title{Impact of Control Strategies on the Control Co-Design of Spar Floating Offshore Wind Turbines
%}

%\newcommand{\titletext}{Modeling and Dynamic Simulation of a Hybrid Wind–Wave System on a Hexagonal Semi-Submersible Platform}
\title{Modeling and Dynamic Simulation of a Hybrid Wind–Wave System on a Hexagonal Semi-Submersible Platform}

\author[umich]{Saeid Bayat\corref{cor1}}
\author[sky]{Jerry Zuo}
\author[umich]{Jing Sun}

\address[umich]{Department of Naval Architecture and Marine Engineering, University of Michigan, Ann Arbor, MI, USA}
\address[sky]{Skyline High School, Ann Arbor, MI, USA}

\cortext[cor1]{Corresponding author. Email: saeidb@umich.edu. Address: 1085 South University, 160 West Hall, Ann Arbor, MI, USA.}

%% Abstract should be no more than 250 words
\begin{abstract}
Offshore renewable energy systems offer promising solutions for sustainable power generation, yet most existing platforms harvest either wind or wave energy in isolation. This study presents a hybrid floating offshore platform that integrates a wind turbine with three oscillating surge wave energy converters (WECs) into a hexagonal semi-submersible structure. In this configuration, the flaps are integrated with the platform geometry to provide both energy extraction and hydrodynamic stability. A modeling and simulation framework was developed using WEC-Sim and benchmarked against the NREL 5~MW semisubmersible reference. Metacentric height analysis confirmed hydrostatic stability across a range of prescribed flap angles. Sensitivity analysis of twelve geometric variables identified flap dimensions and tower length as dominant drivers of stability, energy capture, and tower stress. Time-domain simulations revealed dependence on wave incidence angle, with variations in flap power sharing, capture width ratio (CWR), and platform response. The feasibility of using flap sweeps to modulate pitch motion was also demonstrated. Annual energy production (AEP) estimates based on site-specific data indicate 16.86~GWh from wind and 3.65~GWh from wave energy, with WECs contributing about 18\% of the total. These results highlight the potential of integrated wind–wave platforms and point toward future studies on structural modeling and advanced control.
\end{abstract}


\end{frontmatter}
%\linenumbers

\section{Introduction}
\label{sec:intro}

The rapid growth in global energy demand, along with the environmental consequences of fossil fuels, has intensified the pursuit of sustainable and renewable energy sources. Among the available options, marine renewable energy (MRE), including offshore wind, wave, tidal, and current energy, has gained increasing attention due to the vast and largely untapped potential of the oceans, which cover more than 70\% of the Earth's surface. Harnessing energy from marine environments offers a promising path toward reducing carbon emissions and supporting the global energy transition. In this context, technologies targeting marine energy are being actively explored and developed to meet future energy demands while preserving ecological balance~\cite{dong2022state,bayat2025nested,lee2025multidisciplinary}.

Wind energy, especially in offshore environments, has emerged as one of the most mature and economically viable renewable technologies. Fixed-bottom offshore wind turbines (OWTs), with capacities exceeding 10~MW, have been successfully deployed in various regions~\cite{wikipedia2025vestasV164,mingyang2023myse16,ge2020haliade13mw,siemensgamesa2020sg14-222,vestas2021v236-15mw}, taking advantage of stronger and more consistent wind resources above sea. Compared to land-based wind, offshore wind offers several advantages: wind speeds offshore are typically higher, less turbulent, and more uniform, resulting in improved energy capture efficiency and reduced fatigue loading on turbine components. Additionally, offshore installations reduce land-use conflicts and visual or noise concerns often associated with onshore wind farms~\cite{lee2025wind}. The offshore wind sector has seen continuous growth, supported by reducing production costs and advances in floating platform technologies. According to the Global Wind Energy Council (GWEC), offshore wind installations reached record highs in recent years, with China contributing significantly to global capacity expansion. Globally, about 83~GW of offshore wind capacity had been installed by the end of 2024, with approximately 8~GW added that year alone~\cite{GWEC2025GlobalOffshoreWindReport}. Projections suggest this could rise to nearly 441~GW by 2034 if current growth trends continue~\cite{GWEC2025GlobalOffshoreWindReport}. Given these developments, as the industry moves into deeper waters to access stronger and more consistent wind resources while overcoming the limited availability of shallow sites, the challenge of designing cost-effective and robust support structures has become a key research area.

While fixed-bottom offshore wind turbines have been successfully deployed in shallow and moderate water depths, their applicability becomes limited as depth increases beyond approximately 50--60 meters due to rising installation and structural costs~\cite{barooni2022floating}. To extend offshore wind development into deeper waters, floating offshore wind turbines (FOWTs) have been developed as a technically feasible and increasingly promising solution. Various platform designs---such as spar-buoy, semi-submersible, and tension leg platforms---have been engineered to support large turbine systems while maintaining stability under dynamic marine conditions and accommodating diverse seabed profiles. While the technology is still evolving, recent pilot projects such as Hywind Scotland~\cite{equinor2017hywind} and WindFloat Atlantic~\cite{principlepower2020windfloat} have demonstrated the technical feasibility of FOWTs and have paved the way for larger commercial deployments. As floating wind continues to mature, it is expected to play an increasingly vital role in expanding global offshore wind capacity. By the end of 2024, global floating offshore wind capacity amounted to only about 278 MW~\cite{NortonRoseFulbright2025FloatingWind}. However, the development pipeline is substantial—around 244 GW—and long-term projections indicate that installed capacity could approach 217 GW by 2050~\cite{AOWA2025BeyondTheHorizon}.

In parallel with the growth of offshore wind, wave energy has emerged as a promising yet less mature marine renewable energy source. Ocean waves offer a high energy density and exhibit greater predictability and consistency over longer timescales compared to wind, making them attractive for energy generation~\cite{wisse2018,barbar2017}. Studies have estimated that wave energy could meet a substantial portion of global electricity demand---for instance, over 40\% in the UK and 34\% in the USA~\cite{barbar2017}. Nonetheless, commercial deployment remains limited due to critical challenges—including device survivability in harsh sea conditions, high capital and maintenance costs, and relatively low conversion efficiency compared to wind technologies—factors that have slowed wide-scale adoption~\cite{aderinto2018}. These limitations have catalyzed innovations in WEC design, including the integration of hybrid or co‑located systems that couple wave and wind resources to improve efficiency and economic feasibility~\cite{aderinto2018,mctiernan2020review}.

Combining wind and wave energy in a single system offers an effective solution to overcome the limitations of each technology when used independently. Wind and wave resources are often temporally complementary—waves typically persist longer than wind events—resulting in smoother and more continuous energy production when both are harnessed together~\cite{said2023complementarity}. Hybrid systems, where wind turbines and wave energy converters share a single platform or foundation, can reduce the Levelized Cost of Energy (LCOE) by minimizing infrastructure duplication, optimizing marine space usage, and lowering installation and maintenance costs~\cite{dong2022state}. The hybrid design may also reuse existing infrastructure, such as decommissioned oil and gas platforms, offering further economic benefits~\cite{mctiernan2020review}. Additionally, the integration of WECs can mitigate platform motion and reduce wave loads on floating wind turbines, improving aerodynamic performance and structural lifespan ~\cite{chen2022load,zhang2022coupled}.
%Various control strategies—including active and switching control—have been employed to further suppress platform motion and enhance reliability. 
The combination of resources also increases energy availability and overall capacity factor: wave energy is more persistent and predictable, and its phase-lag with wind leads to smoother and more stable power generation~\cite{gaughan2020assessment}.

In recent years, several research programs and prototype demonstrations have been initiated to explore hybrid wind--wave systems. Notable projects include the MARINA Platform~\cite{kringelum2013grid}, MERMAID~\cite{sakata1994multimedia}, TROPOS~\cite{estrada2016applying}, and H2Ocean~\cite{comyn2022h2ocean} in Europe, as well as pilot installations in China such as the Daguan Island project~\cite{xiong2009study} and the Zhaitang Island platform~\cite{yuan2018assessment}. These systems range from co-located configurations—where wind and wave devices operate independently but within the same marine area—to fully integrated hybrid platforms sharing structural components. Recent studies have shown that optimized hybrid layouts can significantly increase total energy output, reduce fatigue-inducing loads on turbines, and enhance overall system availability. However, hybrid systems also face critical challenges: most are still in conceptual or early prototype stages, and improper integration can increase system complexity and dynamic loading on the structure. Structural concerns such as increased pitch, fatigue risk, and mooring loads under extreme conditions must be carefully addressed. In some cases, hybridization has led to amplified surge motion or increased vertical forces, depending on design and sea states~\cite{zhou2023power}. Additionally, the chosen deployment site may not be optimal for either energy source individually, leading to suboptimal energy capture. To ensure safe and cost-effective deployment, future designs should consider joint wind--wave resource distributions, tailored mooring systems, and coupled simulations. Moreover, a standardized evaluation framework is urgently needed to guide consistent and reliable development. This includes modules for resource assessment, economic modeling, and dynamic analysis to support system optimization and accelerate commercialization.

In this study, we introduce a novel hexagonal floating platform~\cite{OCEANS2025GreatLakes} that combines a central tower supporting a horizontal-axis wind turbine and three flap-type wave energy converters (WECs) mounted on alternating sides of the six-sided semi-submersible structure (see Fig.~\ref{fig:hexagon_view}(a)). Unlike previous designs in which WECs are merely add-ons to floating platforms, the flaps in our system are integral to the platform’s structure. The WECs contribute essential properties---such as buoyancy and hydrodynamic stiffness---without which the platform would lack sufficient stability. As such, the floating wind platform and WEC systems must be co-designed to ensure overall stability and functionality. This integration offers two key advantages: it reduces costs by combining the wave energy conversion function with the structural platform and increases total power output by harvesting energy from waves in addition to wind. Furthermore, by actively controlling the WEC motions, the system can dampen platform dynamics and mitigate undesirable motions. 

The remainder of this paper is organized as follows. Section~\ref{sec:modelling} defines the system geometry and presents the modeling framework, including hydrodynamic, aerodynamic, mooring, and environmental inputs. Section~\ref{sec:rslts} presents the results, where time-domain dynamic simulations are performed to evaluate platform stability, directional wave response, and design sensitivity. Section~\ref{sec:lim_ftr_wrk} discusses key limitations of the current modeling approach and outlines directions for future work. Finally, Section~\ref{sec:cncl} summarizes the main conclusions.

\section{Modeling}
\label{sec:modelling}

\begin{figure}[ht!]
    \centering
    \includegraphics[width=0.75\linewidth]{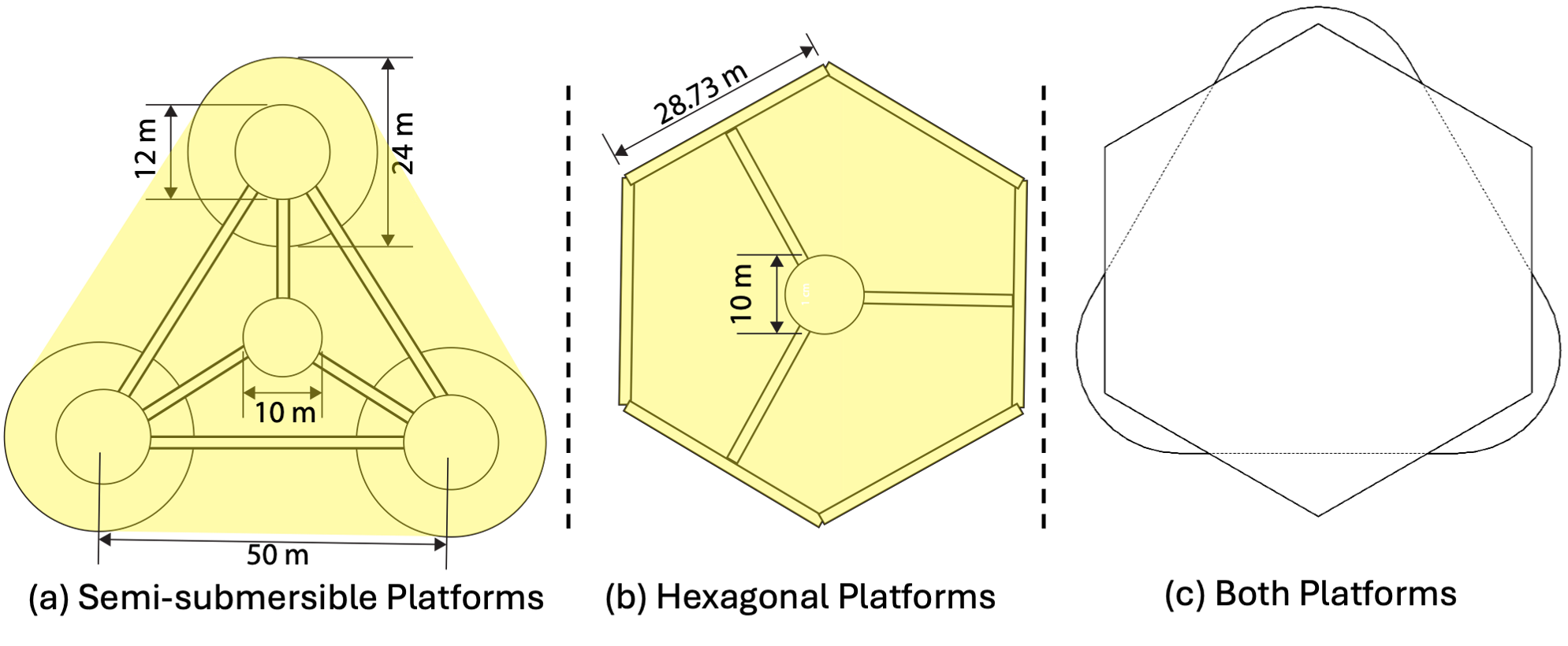}
    \caption{Top views of platform footprints used for scaling. (a) NREL 5~MW semi-submersible reference platform. (b) Proposed hexagonal platform with equivalent footprint area. (c) Overlay comparison of both platforms, shown to the same scale.}
    \label{fig:platfrm_view_dimension}
\end{figure}

To establish a baseline for the scale of the proposed design, the NREL 5~MW semi-submersible platform~\cite{robertson2014definition} is used as a reference, with its dimensions illustrated in Figure~\ref{fig:platfrm_view_dimension}(a). To enable a meaningful comparison, the hexagonal platform is configured to match the footprint area of this reference design, which is approximately \(2{,}145\,\text{m}^2\). For a regular hexagon to enclose an equivalent area, each side must be approximately 28.73~m in length, as shown in Figure~\ref{fig:platfrm_view_dimension}(b). A direct overlay of both platform footprints is provided in Figure~\ref{fig:platfrm_view_dimension}(c), demonstrating that the overall scale of the proposed concept is comparable to that of the reference.

\begin{figure}[ht!]
    \centering
    \includegraphics[width=0.7\linewidth]{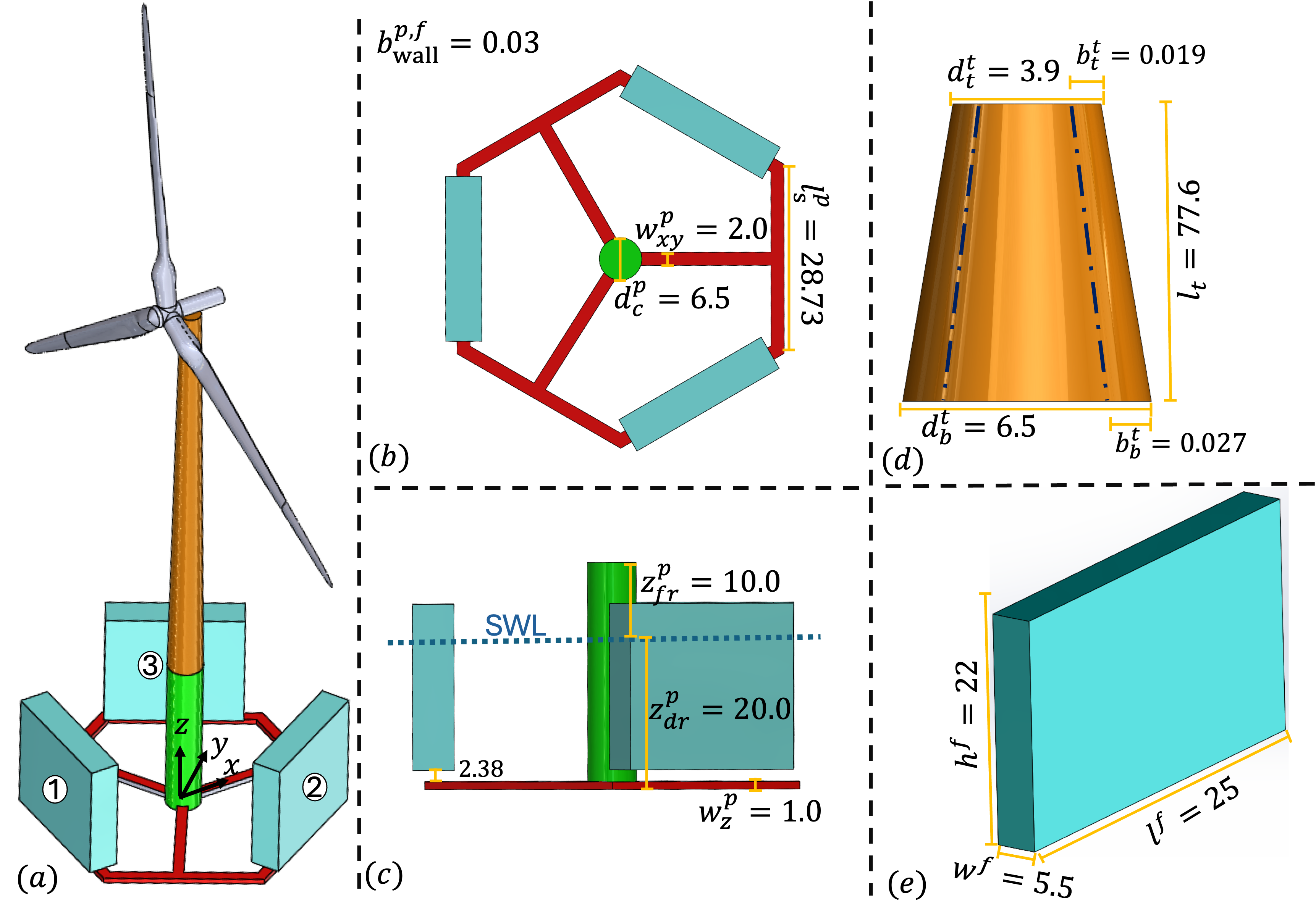}
     \caption{Geometric variables of the hybrid hexagonal platform. 
(a) Full assembly showing the hexagonal platform, central cylinder, wind turbine tower, rotor–nacelle assembly, and three flaps. 
(b) Key geometric variables of the platform, including side length ($l_s^p$), width in the $xy$-plane ($w_{xy}^p$), and central cylinder diameter ($d_c^p$). 
(c) Dimensions in the $xz$-plane, including freeboard height ($z_{fr}^p$), draft ($z_{dr}^p$), and platform height in the vertical direction ($w_z^p$); the flap is mounted 2.38~m above the platform to prevent collision during rotation. 
(d) Tower geometry, including base and top outer diameters ($d_b^t$, $d_t^t$), base and top thicknesses ($b_b^t$, $b_t^t$), and total tower length ($l_t$). 
(e) Flap geometry including width ($w^f$), height ($h^f$), and length ($l^f$). Superscripts $p$, $f$, and $t$ denote platform, flap, and tower, respectively.}
    \label{fig:hexagon_view}
\end{figure}

Figure~\ref{fig:hexagon_view}(a) illustrates the full hybrid energy system, consisting of a hexagonal floating platform with an integrated central cylinder, three flap-type wave energy converters (WECs), a wind turbine tower, and a rotor--nacelle assembly. This hybrid system simultaneously harnesses wind and wave energy: the wind turbine extracts aerodynamic power through its rotor, while the three flaps provide buoyancy and simultaneously serve as WECs that rotate in response to incoming waves. This relative rotation enables the conversion of wave-induced motion into hydrodynamic power, making the flaps essential components for wave energy extraction. In this configuration, Flap~1 is oriented perpendicular to the surge (x) direction, while the other two flaps are positioned at $\pm60^\circ$. Consequently, when waves propagate along the x-axis, Flap~1 generates the highest power output, while the other two contribute less. As the wave direction shifts, the relative energy contribution of each flap changes accordingly.

Figure~\ref{fig:hexagon_view} illustrates the main design variables that define the geometry of the platform, flaps, and tower. The superscripts $p$, $f$, and $t$ correspond to the platform, flap, and tower, respectively. Widths are denoted by $w$, diameters by $d$, thicknesses by $b$, and lengths by $l$. Figure~\ref{fig:hexagon_view}(b) highlights several key geometric parameters, including the platform side length ($l_s^p$), platform width in the $xy$-plane ($w_{xy}^p$), and the diameter of the central cylinder ($d_c^p$). The wall thicknesses of the platform and flaps are denoted by $b_{\mathrm{wall}}^{p,f}$. Figure~\ref{fig:hexagon_view}(c) presents the dimensions in the $xz$-plane, where $z_{fr}^p$ is the freeboard height, defined as the vertical distance from the still water level (SWL) to the deck, $z_{dr}^p$ is the draft, and $w_z^p$ represents the overall height of the platform in the vertical direction. The flap is positioned approximately $2.38$~m above the platform, calculated as $w^{\mathrm{f}}/2 \times \sin(60^\circ)$, to avoid collision when rotated up to $60^\circ$.  Figure~\ref{fig:hexagon_view}(d) shows the tower geometry, including the base and top outer diameters, wall thicknesses, and the total height. Figure~\ref{fig:hexagon_view}(e) details the flap dimensions: width ($w^f$), height ($h^f$), and length ($l^f$).

Among the variables defined in Figure~\ref{fig:hexagon_view}, we select the following as the design variables to be studied in this paper, while the others are kept fixed. 
We divide them into three vectors: the platform vector 
$\mathbf{x}_p = [ z_{\mathrm{dr}}^p,\, z_{\mathrm{fr}}^p,\, l_s^p,\, w_{xy}^p,\, w_z^p,\, d_c^p ]$, 
the flap vector $\mathbf{x}_f = [ l^f,\, h^f,\, w^f ]$, 
and the tower vector $\mathbf{x}_t = [ l_t,\, d_b^t,\, d_t^t ]$. 
The baseline values for these geometric parameters are also indicated in Figure~\ref{fig:hexagon_view}. The tower height, diameter, and wall thicknesses are based on the NREL 5~MW reference wind turbine~\cite{jonkman2009definition}. As discussed previously, the hexagonal platform was designed to have the same footprint area as the NREL semi-submersible platform. The flap dimensions were selected to be on a comparable scale to the platform, providing sufficient buoyancy and contributing to the effective waterplane area and structural stiffness necessary for maintaining system stability—an aspect further examined in subsequent sections.

To perform the dynamic simulation of the hybrid system, several key parameters must first be computed. These include hydrostatic and hydrodynamic coefficients (such as added mass and radiation damping), control parameters for the wave energy converters, wind turbine aerodynamic data, mooring characteristics, and wind turbine control settings. The computation of these inputs is detailed in the following sections.

\subsection{Boundary Element Method for Wave–Body Interaction}
\label{sec:BEM}

\begin{figure}[ht!]
    \centering
    \includegraphics[width=0.4\linewidth]{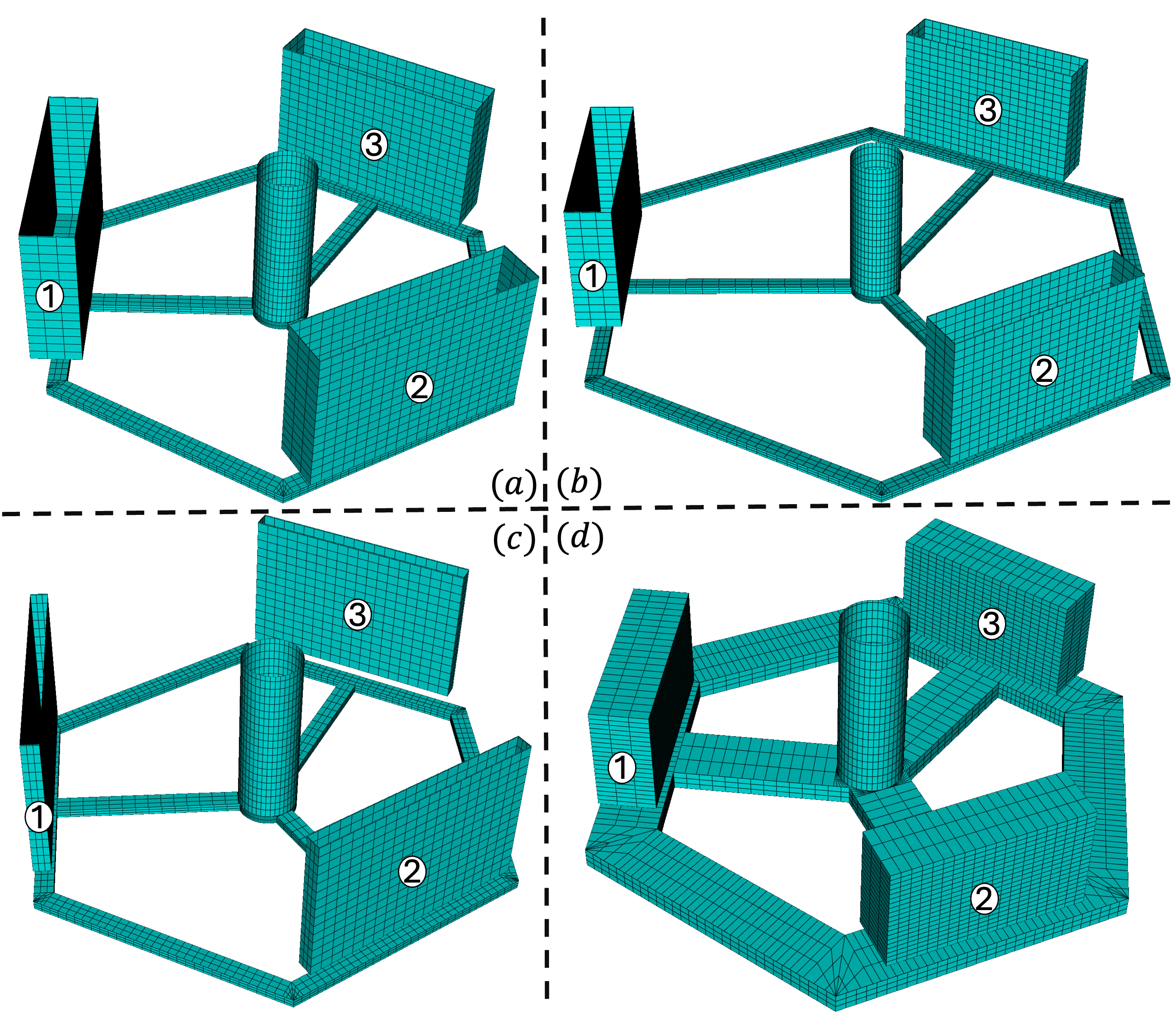}
    \caption{Examples of automatically generated boundary-element method (BEM) meshes for the hybrid platform and flap system under different geometric design configurations. The meshes are created using a Python-based procedure within \texttt{Capytaine}, allowing rapid updates in response to changes in geometric variables from the platform vector $\mathbf{x}_p$ (e.g., platform diameter) and the flap vector $\mathbf{x}_f$ (e.g., flap length), and include only the submerged portions of the structures, which are relevant for hydrodynamic analysis.}
    \label{fig:BEM_diff_dsgns}
\end{figure}

To compute the hydrodynamic and hydrostatic coefficients of the hybrid platform, we use a Boundary-Element Method (BEM) solver—\texttt{Capytaine}~\cite{ancellin_capytaine_2019}, an open-source code widely used for wave–structure interaction analysis. In this regard, the first step involves generating an appropriate mesh of the platform and flaps. In general, this can be done in two ways: (1) by designing the geometry in CAD software (e.g., SolidWorks), exporting the CAD file, and using a meshing tool like \texttt{Gmsh}~\cite{geuzaine2009gmsh} before importing it into \texttt{Capytaine}, or (2) by generating the mesh directly in \texttt{Capytaine} using Python code by defining vertices and planes based on geometric parameters. In this study, we adopt the second approach, which enables automatic mesh generation driven by parametric design variables. This method significantly streamlines design optimization and design-of-study workflows, as mesh updates are triggered automatically when geometry variables change. In contrast, the CAD-based approach requires manual updates each time the geometry changes: the CAD model must be revised, the mesh regenerated in a separate tool, and the result imported into \texttt{Capytaine}. This process is time-consuming and prone to error, making it impractical for parametric studies. To demonstrate the flexibility of our approach, Fig.~\ref{fig:BEM_diff_dsgns} shows example meshes generated for different platform and flap geometries, each created in under one second. Note that the meshes shown represent only the submerged portions of the geometry.

In addition to geometric variables, the inputs to \texttt{Capytaine} must include the masses and centers of gravity of the platform and flaps in order to compute the metacentric height ($GM$). Unlike conventional floating platforms such as spar buoys or semi-submersibles—where the structure has six degrees of freedom (DOF) and a fixed structural geometry during operation—the proposed hybrid system includes rotating flaps. As each flap rotates about its pitch axis, the platform effectively gains three additional DOFs, and the overall system geometry changes dynamically based on the flap angles. This structural variability directly influences the metacentric height, which can become negative for certain configurations. For example, when one flap rotates to $-60^\circ$ and another to $+60^\circ$, the system may become unstable due to an unfavorable shift in hydrostatic properties. Therefore, it is essential to ensure that the metacentric height remains positive for all allowable platform and flap rotation angles. If a given set of geometric parameters results in a negative metacentric height under any allowable flap and platform rotations, the design is considered unstable and is excluded from further analysis in both the BEM solver and system dynamic solver (\texttt{WEC-Sim}~\cite{wecsim}), as both tools are computationally expensive to run.

\begin{figure}[ht!]
    \centering
    \includegraphics[width=0.6\linewidth]{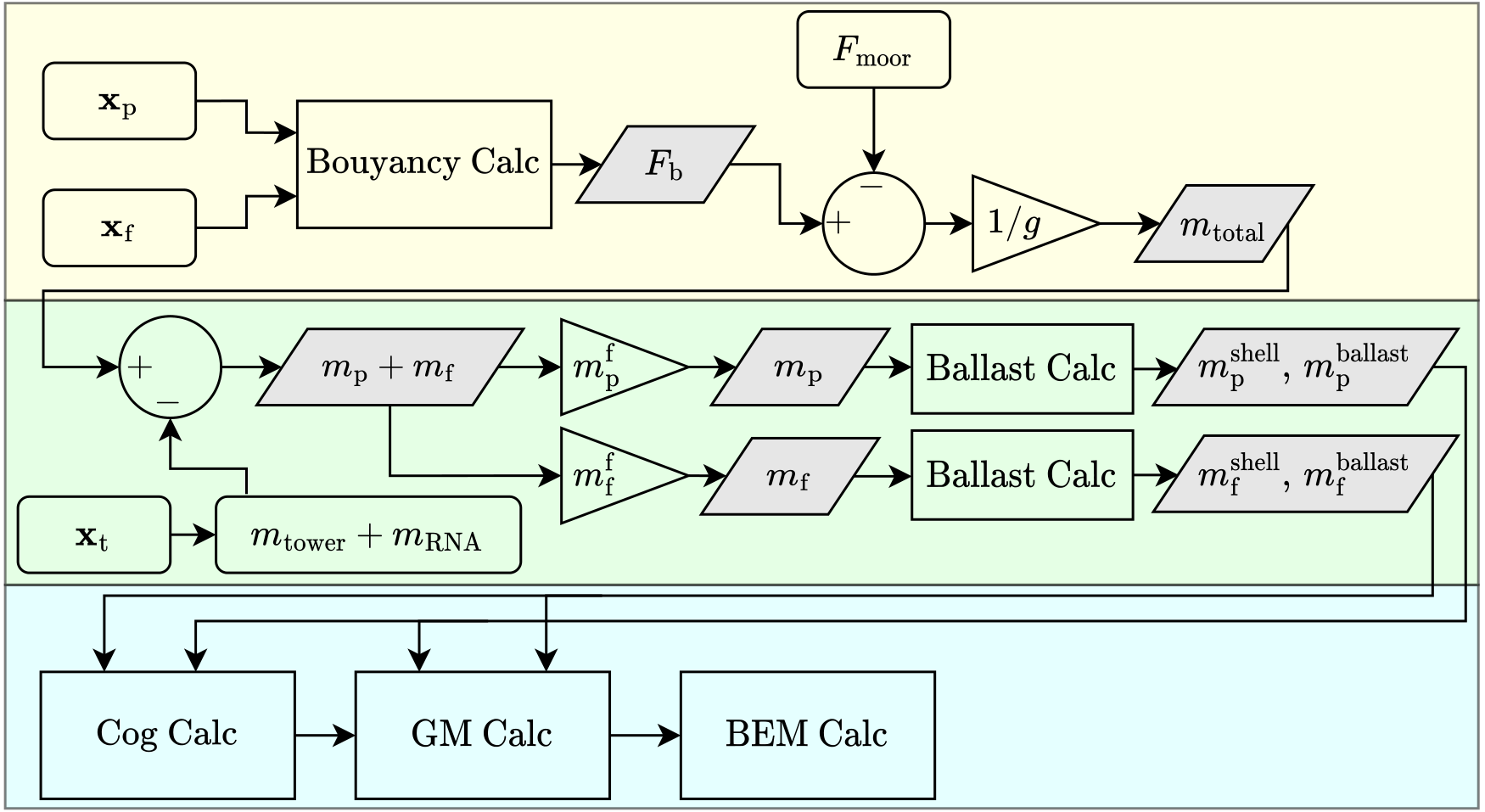}
        \caption{Flowchart for the buoyancy and mass allocation procedure used before the BEM solver. The platform ($x_p$), flap ($x_f$), and tower ($x_t$) design vectors are inputs that determine buoyancy balance, total required mass, and its distribution between platform and flaps, while ensuring the metacentric height ($GM$) remains positive.}
    \label{fig:Hexagon_Calc}
\end{figure}

Figure~\ref{fig:Hexagon_Calc} shows the flowchart for running the BEM computations. Given the platform and flap design variables, denoted as \( \mathbf{x}_p \) and \( \mathbf{x}_f \) vectors, a \texttt{MATLAB} script is used to calculate the buoyancy force. By subtracting the mooring force (will be discussed later) and dividing the result by gravitational acceleration \( g \), the total mass required to balance buoyancy and mooring is determined. Next, using the tower design vector \( x_t \), the mass of the tower and rotor--nacelle assembly is calculated and subtracted from the total required mass. The remaining mass must then be distributed between the platform and the flaps. Two mass fractions are introduced: the platform mass fraction \( m_p^f \) and the flap mass fraction \( m_f^f \), where \( m_p^f + m_f^f = 1 \). These fractions determine how the remaining mass is assigned.

\begin{figure}[ht!]
    \centering
    \includegraphics[width=0.75\linewidth]{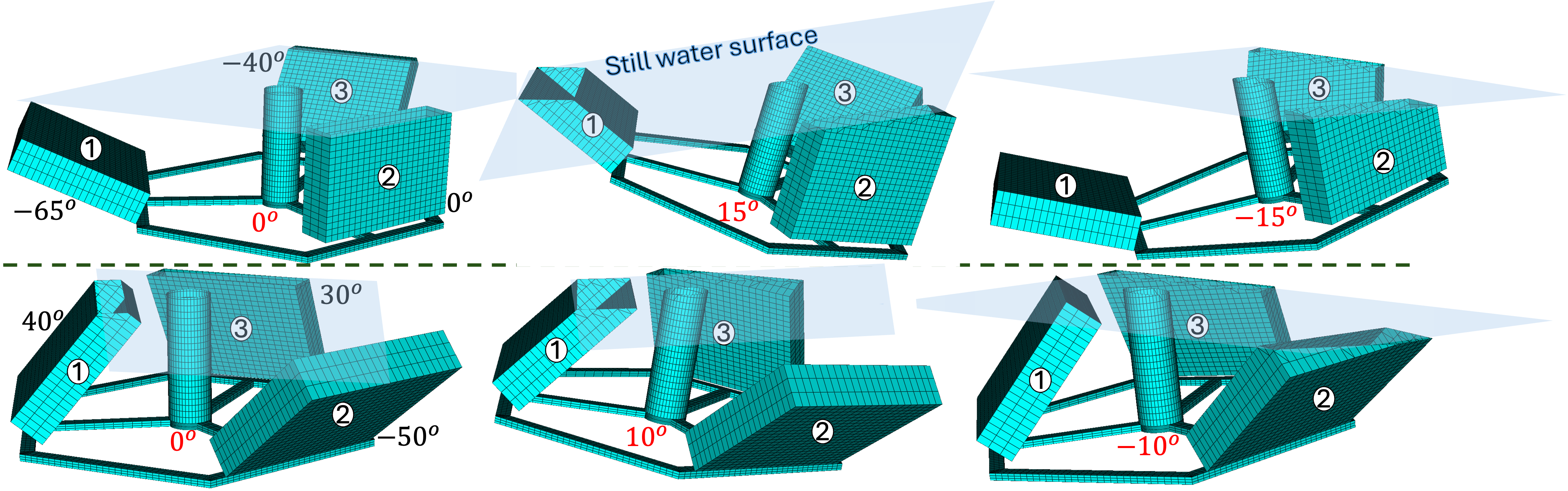}
    \caption{Examples of flap and platform rotation configurations used to compute metacentric heights and evaluate hydrostatic stability. The top row shows configurations with flap angles of $-65^\circ$, $0^\circ$, and $-40^\circ$, while the platform pitch varies from $0^\circ$ to $15^\circ$ and $-15^\circ$. The bottom row shows flap angles of $40^\circ$, $-50^\circ$, and $30^\circ$, with platform pitch angles of $0^\circ$, $10^\circ$, and $-10^\circ$. Each image illustrates the submerged portion of the hybrid platform under distinct combinations of flap and platform rotations.}
    \label{fig:rotations_BEM}
\end{figure}

Then, using the given geometry and wall thicknesses of the platform and flaps, the mass of the steel shell is computed. The required remaining mass is satisfied with ballast, which may consist of slurry ($\rho = 5000~\mathrm{kg/m^3}$), seawater ($\rho = 1025~\mathrm{kg/m^3}$), or a mixture of both. Any unused portion of the internal volume is assumed to be empty (air-filled), which contributes only to buoyancy but not to structural mass. The center of gravity (COG) of the entire system is then calculated. In the next step, meshes of the hybrid system are generated for different platform and flap angle configurations, and the metacentric heights are evaluated using \texttt{Capytaine}. For example, Figure~\ref{fig:rotations_BEM} illustrates the submerged portions of six different platform and flap configurations, each defined by distinct flap and platform angles. If the metacentric height remains positive for all allowable platform and flap rotations (will be discussed later) under a given geometry, the BEM solver proceeds to compute the hydrodynamic and hydrostatic coefficients for that flap and platform geometry ($x_f,\, x_p$).

\begin{figure}[ht!]
    \centering
    \includegraphics[width=0.8\linewidth]{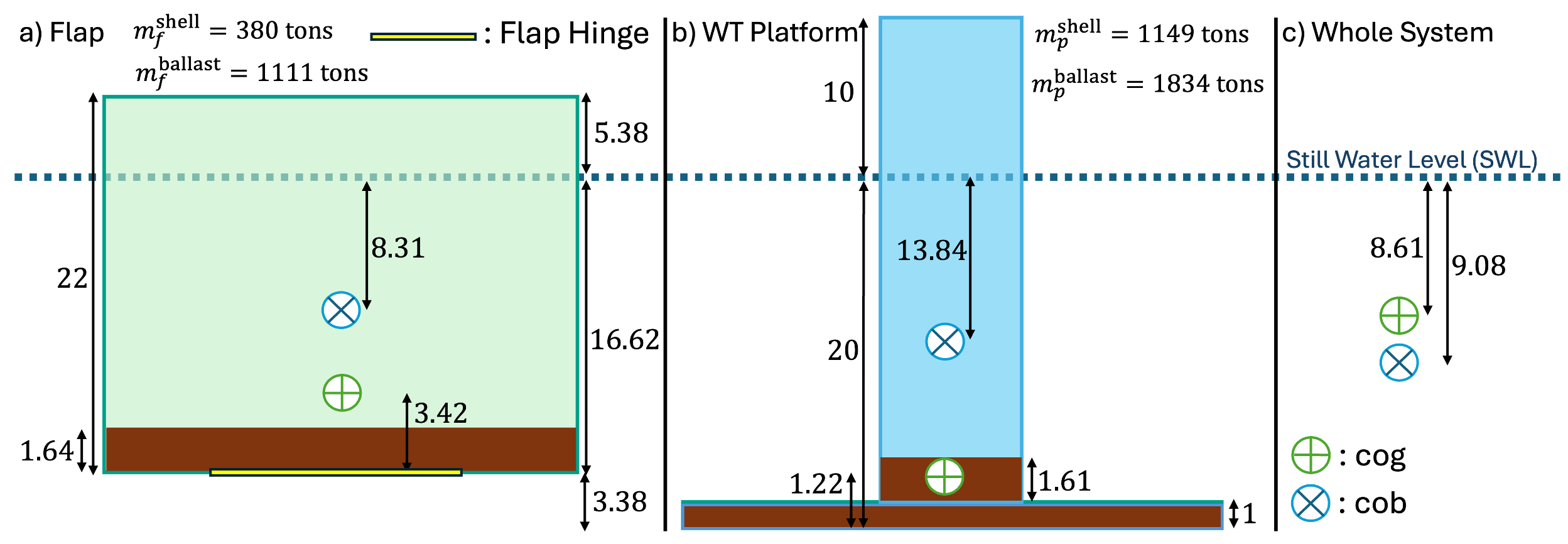}
    \caption{Center of gravity (COG), center of buoyancy (COB), mass distribution, and principal dimensions of the hybrid system. (a) Individual flap configuration, (b) platform without flaps, and (c) full assembly including the tower and rotor--nacelle assembly (RNA). The brown region in panels~(a) and~(b) represents filler material (e.g., slurry) used inside the flap. The full system's COG lies above the COB, emphasizing the need for positive metacentric height (GM) to ensure hydrostatic stability.}
    \label{fig:cog_cob_properties}
\end{figure}

The center of gravity (COG), center of buoyancy (COB), mass distribution, and key geometric dimensions for the baseline hybrid system are shown in Fig.~\ref{fig:cog_cob_properties}. Panel~(a) illustrates the properties of an individual flap, while panel~(b) presents the platform without the flaps, and panel~(c) shows the complete system including the tower and rotor--nacelle assembly (RNA). The material used to fill the flap---such as slurry---is shown in brown in panels~(a) and~(b). Although not shown in this figure, the COG of the tower is located 43.34\,m above the still water level (SWL), with a mass of 249\,tons, while the COG of the RNA is located 87.6\,m above SWL with a mass of 350\,tons. As shown in panel~(c), the COG of the assembled system lies above the COB. This arrangement necessitates verification that the metacentric height (GM) remains positive under all permissible flap and platform rotations. Unlike some conventional platforms---such as spar-type configurations---where the COG naturally lies below the COB, inherently ensuring static stability, the proposed hybrid system does not satisfy this sufficient condition for stability, and thus requires explicit GM evaluation prior to running expensive hydrodynamic simulations.

\begin{figure}[ht!]
  \centering
  % first subfigure
  \begin{subfigure}[b]{0.40\linewidth}
    \centering
    \includegraphics[width=\linewidth]{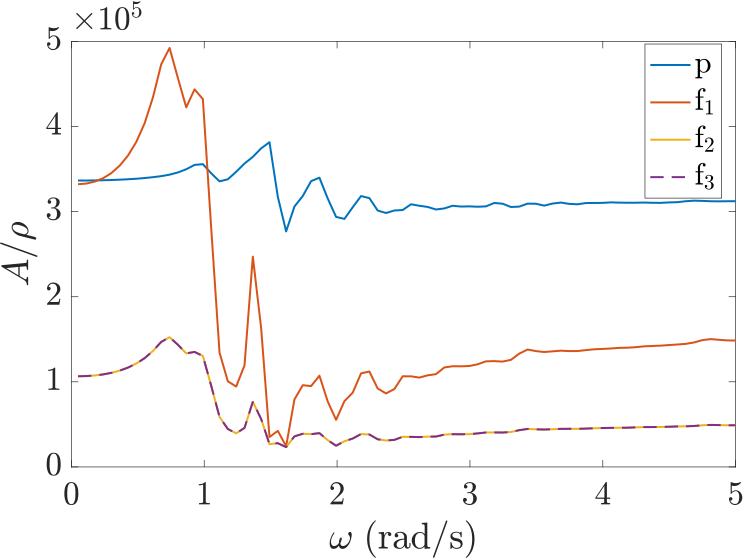}
    \caption{Added mass.}
    \label{fig:sub1}
  \end{subfigure}
  %\hfill
  % second subfigure
  \begin{subfigure}[b]{0.40\linewidth}
    \centering
    \includegraphics[width=\linewidth]{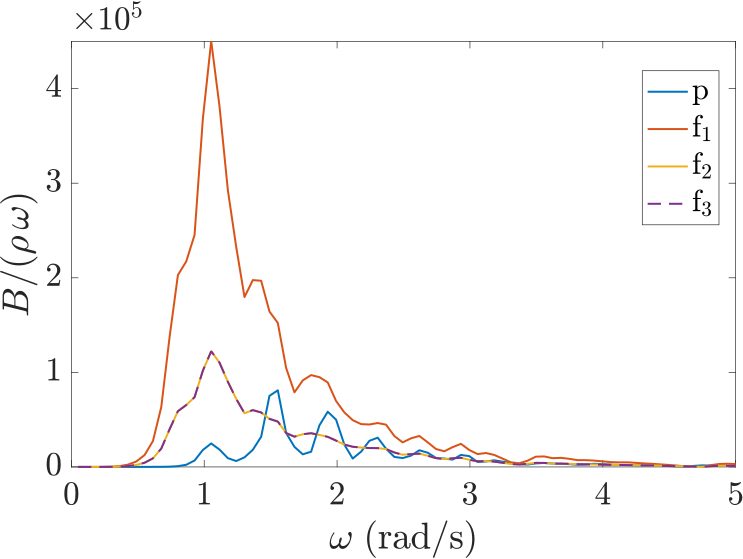}
    \caption{Radiation damping.}
    \label{fig:sub2}
  \end{subfigure}
  \caption{Hydrodynamic coefficients for the pitch DOF of the hybrid WEC–platform system.  
\(p\) denotes the platform; \(f_1\), \(f_2\), and \(f_3\) refer to the three individual flaps.}
  \label{fig:two_plots_BEM}
\end{figure}

An example of the hydrodynamic coefficients for the hybrid system is shown in Fig.~\ref{fig:two_plots_BEM}. 
Figure~\ref{fig:two_plots_BEM}(a) presents the normalized added mass, $A/\rho$, 
while Fig.~\ref{fig:two_plots_BEM}(b) shows the normalized radiation damping, $B/(\rho\omega)$, 
for the platform and the three flap-type wave energy converters (WECs) in the pitch degree of freedom (DOF), 
both plotted as functions of the wave frequency $\omega$. 
Here, $A$ denotes the added mass, $B$ the radiation damping, $\rho$ the water density, 
and $\omega$ the wave frequency. 
As expected, the added mass approaches a constant value at high frequencies, 
whereas the radiation damping tends toward zero. 
In this case, the wave propagation direction is set to $0^\circ$, aligned with the platform’s surge direction. 
Consequently, flap~1—normal to the incoming wave—exhibits a distinct hydrodynamic response, 
while flaps~2 and~3, located symmetrically at $\pm 60^\circ$, display identical behavior, 
as reflected in their overlapping BEM results.

\subsection{WEC PTO Force Modeling}
\label{WEC_PTO}

In this study, the power take-off (PTO) system of the Wave Energy Converter (WEC) employs a passive control strategy. Passive control is modeled as a linear damper that resists motion proportionally to the relative angular velocity between the flap and the platform. The PTO force and the corresponding optimal damping coefficient under regular wave conditions are defined in Eq.~\eqref{eq:PTO_force}, where \(K_p\) is the damping coefficient, and \(\dot{X}\) is the relative pitch angular velocity of the flap with respect to the platform. In Eq.~\eqref{eq:PTO_force}, \(\omega = 2\pi/T\) is the wave frequency, \(T\) is the wave period, \(B(\omega)\) is the frequency-dependent radiation damping, \(A(\omega)\) is the added mass, \(I\) is the inertia of the flap, and \(K_{hs}\) is the hydrostatic restoring stiffness due to buoyancy and gravity~\cite{coe2021practical}.

\begin{equation}
\begin{aligned}
    F_{\mathrm{PTO}} &= -K_p \dot{X}, \\
    K_{p,\mathrm{opt}} &= \sqrt{B(\omega)^2 + \left( \frac{K_{hs}}{\omega} - \omega (I + A(\omega)) \right)^2}.
\end{aligned}
\label{eq:PTO_force}
\end{equation}

Passive damping is widely used in early-stage WEC design due to its simplicity, low implementation cost, and unidirectional power flow—i.e., it only extracts power from the system without requiring external power injection. Although passive control is suboptimal in irregular sea states, it performs reasonably well in regular waves when tuned appropriately using Eq.~\eqref{eq:PTO_force}. While more advanced control strategies—such as PI (proportional-integral) control—can enhance energy capture by introducing a reactive component, they also require bidirectional power flow, meaning energy is at times injected into the system (e.g., via an active motor) to optimize phase alignment with wave excitation forces. Moreover, PI control can overpredict power absorption if PTO efficiency is not carefully modeled, and may lead to physically infeasible motions due to the lack of motion constraints.

Given these challenges and the objective of developing a reliable, low-complexity baseline model for the hybrid wind–wave system, passive damping is adopted in this work. It provides a conservative and robust estimate of power performance, avoiding overprediction that can arise from idealized or overly aggressive control strategies. Additionally, passive control avoids excessive flap motion amplitudes that may result from reactive control methods lacking physical motion constraints. Future studies may explore active and reactive PTO strategies with motion limits to improve performance under irregular wave environments.

It is important to note, however, that the optimal damping coefficient defined in Eq.~\eqref{eq:PTO_force} is derived under the assumption of a single-degree-of-freedom (SDOF) WEC with a fixed bottom. In the hybrid floating system considered here, the flaps are coupled not only through hydrodynamic interactions but also via the shared floating platform, resulting in a multi-degree-of-freedom (MDOF) system. Therefore, the damping coefficients computed using Eq.~\eqref{eq:PTO_force} are no longer truly optimal, but are instead used as approximate values for early-stage design and modeling convenience. Future work will incorporate hydrodynamic coupling and full MDOF dynamics to refine these parameters and evaluate advanced control strategies more accurately under realistic wave conditions.

\subsection{Wind Turbine Controller, Aero-loads, and Mooring.}
\label{sec:win_turbine}

To accurately simulate the hybrid wind--wave system, it is essential to model the aerodynamic loading, wind turbine control, and mooring dynamics in conjunction with the \texttt{WEC-Sim} framework~\cite{wecsim}, a MATLAB-based open-source tool for time-domain dynamic simulation of wave energy converters. In this work, the MATLAB-based Offshore Simulation Tool (MOST)~\cite{sirigu2022development} is employed to enable dynamic co-simulation of floating offshore systems by integrating high-fidelity aerodynamic models, turbine structural dynamics, and advanced control strategies. Within MOST, the mooring system is represented as a quasi-static, nonlinear catenary configuration. Mooring forces and moments are obtained from precomputed look-up tables that relate platform displacements in six degrees of freedom---surge, sway, heave, roll, pitch, and yaw---to the corresponding restoring forces and torques. This approach captures the nonlinear stiffness characteristics of the mooring system while maintaining efficient simulation performance. 

As the baseline geometry shown in Figure~\ref{fig:hexagon_view} differs from that of the NREL 5~MW semi-submersible platform, the resulting buoyancy force is also different, and consequently, the mooring force must be adjusted. The total buoyancy force ($F_b$) of the NREL semi-submersible platform is approximately \(1.4 \times 10^8~\text{N}\), whereas the buoyancy force of the proposed baseline hexagonal hybrid platform (including flaps) is approximately \(0.8 \times 10^8~\text{N}\). This indicates that the submerged volume of the hybrid design is about 57\% of that of the reference semi-submersible platform. Also, the total mooring force at equilibrium for the NREL semi-submersible platform is \(1.84 \times 10^6~\text{N}\).  

To determine a consistent target mooring force for the hexagonal platform, we scaled the semi-submersible mooring force proportionally to the ratio of buoyancy forces, as shown in Eq.~\eqref{eq:f_bouy}. To achieve this target, a scaling factor $s$ (applied to the baseline mooring line diameter) was defined as the design variable in an optimization problem. The objective function minimized the squared error between the target buoyancy force and the restoring force calculated by MOST~\cite{sirigu2022development}. Solving this optimization yielded $s = 1.339$, which was then used to scale the mooring line diameter for the hexagonal platform.

\begin{equation}
    F_{\mathrm{moor}}^{\mathrm{Hex}} = \frac{F_{\mathrm{b}}^{\mathrm{Hex}}}{F_{\mathrm{b}}^{\mathrm{semi\text{-}sub}}} \times F_{\mathrm{moor}}^{\mathrm{semi\text{-}sub}} = 1.05 \times 10^6~\text{N}.
    \label{eq:f_bouy}
\end{equation}

In addition to the mooring, accurate modeling of aerodynamic forces is essential for simulating the hybrid wind--wave system. Aerodynamic forces acting on the rotor are computed using blade element momentum theory (BEMT), which resolves the response of each blade based on local wind speed, rotor speed, and blade pitch angle. During simulation, these forces are interpolated from precomputed look-up tables that span a range of operating conditions. The model also accounts for the influence of platform motion on the apparent wind experienced at the rotor by incorporating effects such as hub surge, pitch velocity, and yaw rotation. This dynamic coupling allows for a realistic interaction between the aerodynamic loads and the floating platform.

Furthermore, the simulation incorporates realistic wind inputs and turbine control strategies to capture the coupled dynamics of the hybrid wind--wave system.
 The wind speed is resolved along the blade span to account for radial inflow variations, improving the accuracy of aerodynamic force predictions. Wind turbine control is implemented using the Reference Open-Source Controller (ROSCO)~\cite{ROSCO_toolbox_2021}, which governs generator torque and collective blade pitch based on operational regions. Below rated wind speed, the controller modulates generator torque to maximize aerodynamic efficiency. Above rated wind speed, blade pitch is adjusted to regulate rotor speed while generator torque remains fixed. This approach maintains power output within safe operational limits and mitigates mechanical loads under high wind conditions~\cite{bayat2025impact}.

In this study, controller parameters are fixed. The generator torque controller is tuned with a natural frequency of \(\omega_{\mathrm{g}}^\mathrm{n} = 0.12~\mathrm{rad/s}\) and a damping ratio of \(\xi_{\mathrm{g}} = 1.5\). The blade pitch controller has a natural frequency of \(\omega_{\theta}^\mathrm{n} = 0.33~\mathrm{rad/s}\) and a damping ratio of \(\xi_{\theta} = 5.1\). To improve platform stability, a floating feedback term is added to the blade pitch control using a proportional gain of \(K_v = 0.31\). This term modifies the blade pitch in response to platform pitch motion, helping to reduce excessive oscillations. Overall, this integrated modeling approach captures the dynamic coupling between the wind turbine, platform, and mooring system, enabling realistic performance evaluation of hybrid offshore energy devices in dynamic environmental conditions.

\subsection{Wind $\&$ Wave data}
\label{sec_wind_wave}

\begin{figure}[ht!]
    \centering
    \includegraphics[width=0.75\linewidth]{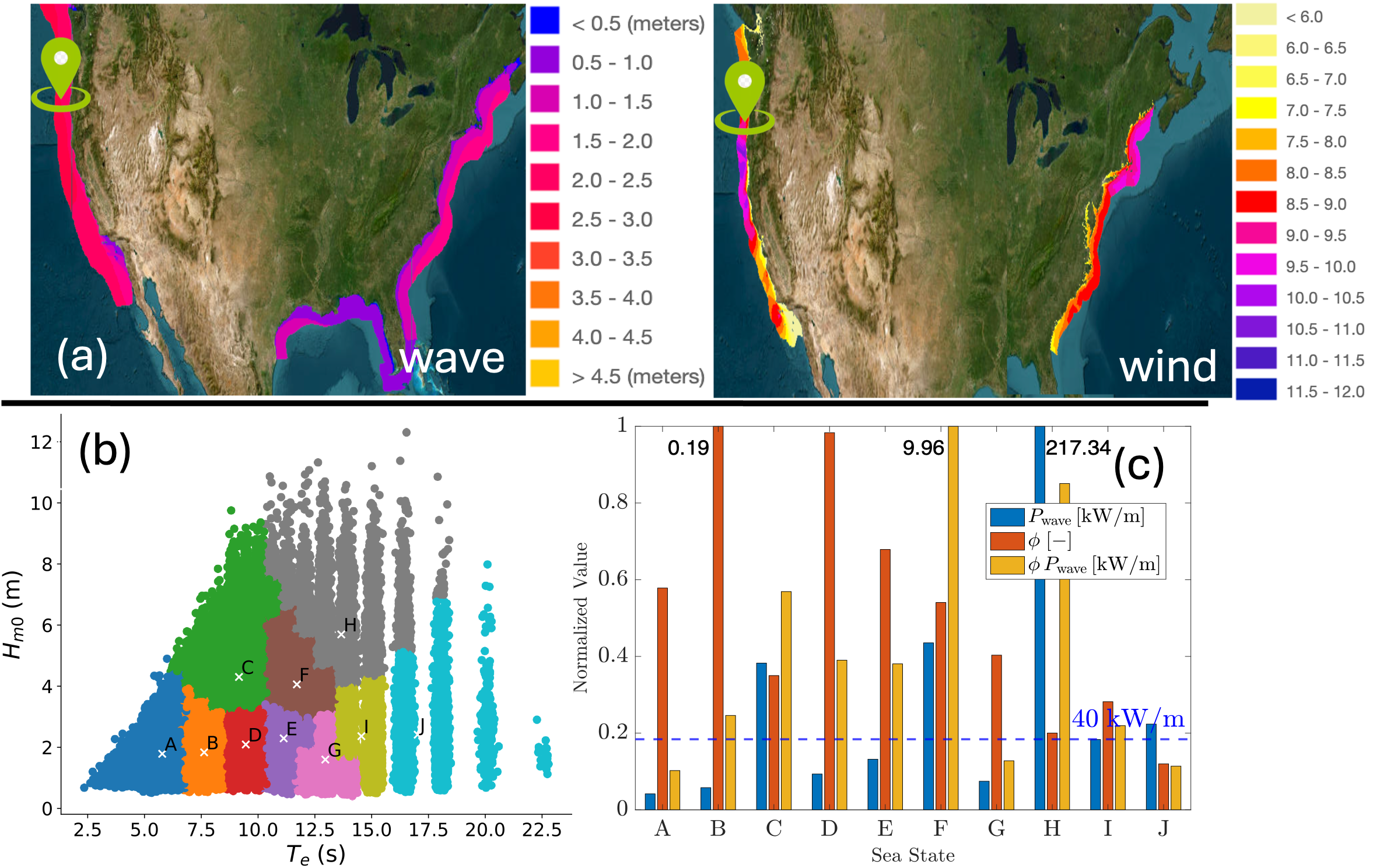}
    \caption{(a) Significant wave height along the U.S. shoreline, highlighting greater wave energy potential on the West Coast~\cite{databasin_swh_annual}. (b) Historical wave data (2005–2024) from CDIP Station 139 near the Oregon coast, plotted by significant wave height (y-axis) and energy period (x-axis). Data are clustered into 10 sea states (A to J) using $k$-means clustering. (c) Normalized comparison of wave power density, weight of occurrence, and their product (weighted power) for Sea States A to J. Actual maximum values used for normalization are displayed beside each column.} %Sea State~D shows the highest weighted power and is selected for further analysis.}
    \label{fig:wave_data}
\end{figure}

Selecting an appropriate location for deploying the hybrid wind--wave energy system is crucial, as the availability of wind speed and wave power directly affects the energy generation performance of both the wind turbine and the wave energy converter. Therefore, the chosen site must offer consistently strong wind and wave resources. As shown in Figure~\ref{fig:wave_data}(a), the West Coast of the United States generally experiences higher wind speeds and significant wave heights (and longer wave periods, although the latter are not shown in the figure), which together contribute to a higher wave power density. Consequently, the Oregon coast is selected in this study as a suitable potential location for hybrid device deployment.

Figure~\ref{fig:wave_data}(b) shows wave data obtained from the Coastal Data Information Program (CDIP) station 139, located near the west coast of Oregon. In this figure, the x-axis represents the energy period (in seconds), and the y-axis indicates the significant wave height (in meters). The plot displays historical wave data collected between 2008 and 2024, with each data point representing a 30-minute interval. After gathering this dataset, we applied $k$-means clustering to categorize the data into 10 distinct clusters. The centroids of each cluster are marked using white symbols and are labeled as Sea States A to J.  Figure~\ref{fig:wave_data}(c) shows the wave power density (\(P_{\mathrm{wave}}\)), the probability of occurrence (\(\phi\)), and their product—the weighted power (\(\phi P_{\mathrm{wave}}\))—for Sea States A through J. The wave power density is calculated using Eq.~\ref{eq:P_wave}, where \(\rho\) denotes the water density, \(g\) is the gravitational acceleration, \(H_{m_0}\) is the significant wave height, and \(T_e\) is the energy period.

Figure~\ref{fig:wave_data}(c) presents the occurrence probability, wave power density, and weighted power contribution for ten representative sea states labeled A through J. Each dataset is normalized by its respective maximum value, allowing for direct visual comparison across categories. The original (non-normalized) maximum values used for scaling are indicated next to the corresponding bars. Based on this weighted power distribution, the average wave power density is approximately 40~kW/m. For the subsequent analysis, we adopt a representative JONSWAP wave spectrum with a significant wave height of 2.85~m and an energy period of 10.04~s~\cite{dunkle2020pacwave}, corresponding to the average wave power density of 40~kW/m.

\begin{equation}
    P_{\mathrm{wave}} = \frac{\rho g^2}{64\pi} H_{m_0}^2 T_e\,\, , \,\, \mathrm{[W/m]}.
    \label{eq:P_wave}
\end{equation}

%As shown, Sea State~D has the highest weighted power, meaning it contributes most significantly to wave energy generation at the selected location (the west coast of Oregon). Therefore, Sea State~D—characterized by a significant wave height of 3.95~m and an energy period of 9.67~s—is used for further analysis in the following sections. 

Building on the environmental site selection discussed above, it is equally important to evaluate the wind turbine under realistic operating conditions to ensure robust performance and structural integrity. While full-system design studies typically cover a broad set of Design Load Cases (DLCs), early-stage analysis often focuses on a smaller subset due to the high computational cost of dynamic simulations. In this study, a 600-second turbulent wind profile with a mean wind speed of 11.35\,m/s is used. This wind speed corresponds to the transition between Region~2 and Region~3 of the turbine’s power curve, where the thrust force peaks. This transitional region is known to impose the highest combined aerodynamic and structural loads, making it a critical scenario for evaluating system response~\cite{lee2025wind}. The selected mean wind speed is defined at the turbine’s hub height of 77.6\,m, aligning with the tower configuration considered here.

\section{Results}
\label{sec:rslts}

This section presents the results of the proposed hybrid wind--wave energy platform, organized into five parts. In Subsection~\ref{subsec:mtch}, the hydrostatic stability of the platform is assessed through a meta-centric height analysis across various flap and platform configurations. Subsection~\ref{subsec:DOE} introduces a sensitivity analysis, where key geometric parameters are varied to evaluate their influence on platform motion and power output. In Subsection~\ref{subsec:Time}, time-domain simulations are conducted under two wave incidence directions to examine the directional response of the system. Subsection~\ref{subsec:flp_cntrl} explores the use of flap rotation control strategies to improve platform stability. Finally, in Subsection~\ref{subsec:wec_wt}, the annual energy production (AEP) of both the wind and wave subsystems is estimated using site-specific environmental data, demonstrating the potential benefits of hybrid co-design.

\subsection{Metacentric Height and Stability Analysis}
\label{subsec:mtch}

Before performing the computationally expensive BEM calculations and time-domain simulations, 
it is crucial to first verify that the hybrid platform design is hydrostatically stable. That is, the system should remain stable under rotational motions of the flaps and the platform itself. One effective measure of this is the metacentric height, which depends on both the submerged geometry of the system (affecting buoyancy) and the center of gravity (CoG) of the entire structure.

\begin{figure}[ht!]
  \centering
  % first subfigure
  \begin{subfigure}[b]{0.49\linewidth}
    \centering
    \includegraphics[width=\linewidth]{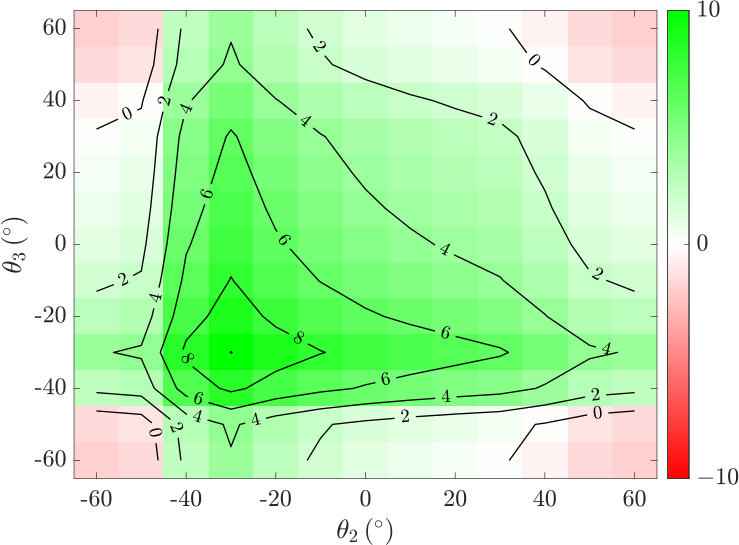}
    \caption{}
    \label{fig:sub1}
  \end{subfigure}
  \hfill
  % second subfigure
  \begin{subfigure}[b]{0.49\linewidth}
    \centering
    \includegraphics[width=\linewidth]{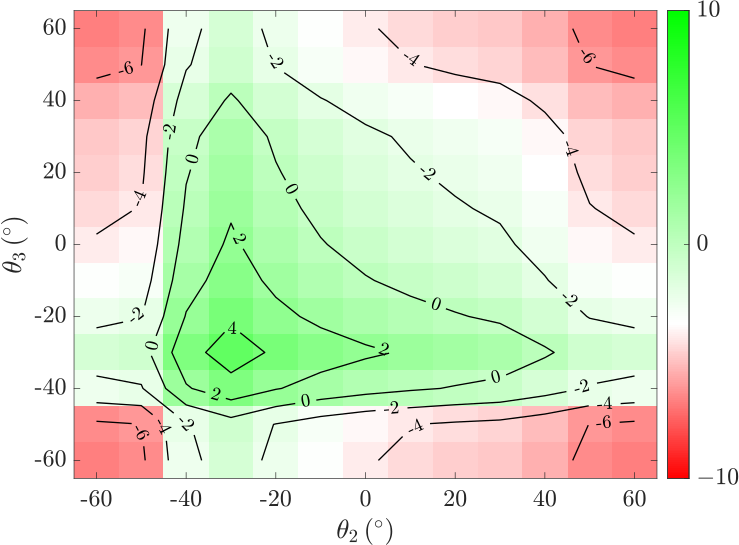}
    \caption{}
    \label{fig:sub2}
  \end{subfigure}
  \caption{Metacentric height as a function of flap rotation for the baseline design. Flap~1 is fixed at $60^\circ$, while Flaps~2 and 3 vary their angles between $-60^\circ$ and $+60^\circ$. (a) shows results using slurry as ballast material; (b) shows results using water. Positive values of metacentric height indicate hydrostatic stability, while negative values indicate instability.}
  \label{fig:two_plots}
\end{figure}

As illustrated in Fig.~\ref{fig:Hexagon_Calc}, the location of the CoG is influenced by the mass distribution, particularly the placement and density of ballast material within the platform and flaps. For example, using a higher-density material such as slurry (relative to water) lowers the CoG, potentially improving stability. To examine this, two ballast cases are considered: one using slurry and the other using water. Figure~\ref{fig:two_plots} presents the resulting metacentric heights for the baseline design configuration. In this setup, the left flap (flap 1) is fixed at $60^\circ$, while the remaining two flaps vary their rotational angles between $-60^\circ$ and $60^\circ$. In this figure, the $x$- and $y$-axes of the plots represent the flap rotation angles, and the color bar indicates the metacentric height.

Figure~\ref{fig:two_plots}(a) shows the results for the slurry ballast case, where the system 
remains stable (i.e., positive metacentric height) for flap rotations between $-40^{\circ}$ and 
$+40^{\circ}$. In contrast, Fig.~\ref{fig:two_plots}(b) corresponds to the water ballast case, 
which exhibits a larger region of instability (negative metacentric height). As observed, the system 
is stable only within approximately $\pm 30^{\circ}$ of flap rotation. These findings indicate that, if water ballast is to be used, the design may require additional modifications—such as increasing flap width to enlarge the waterplane area—or mechanical stops to restrict flap rotation and preserve stability. In this paper, we proceed with the slurry ballast case for all simulations and performance analyses due to its better hydrostatic stability. %However, it is worth noting that the water ballast case offers certain advantages, such as increased system inertia, which could help reduce platform pitch motion. 
%Future studies are recommended to further investigate the performance trade-offs between these two ballast strategies.

\begin{figure}[ht!]
    \centering
    \includegraphics[width=1.0\linewidth]{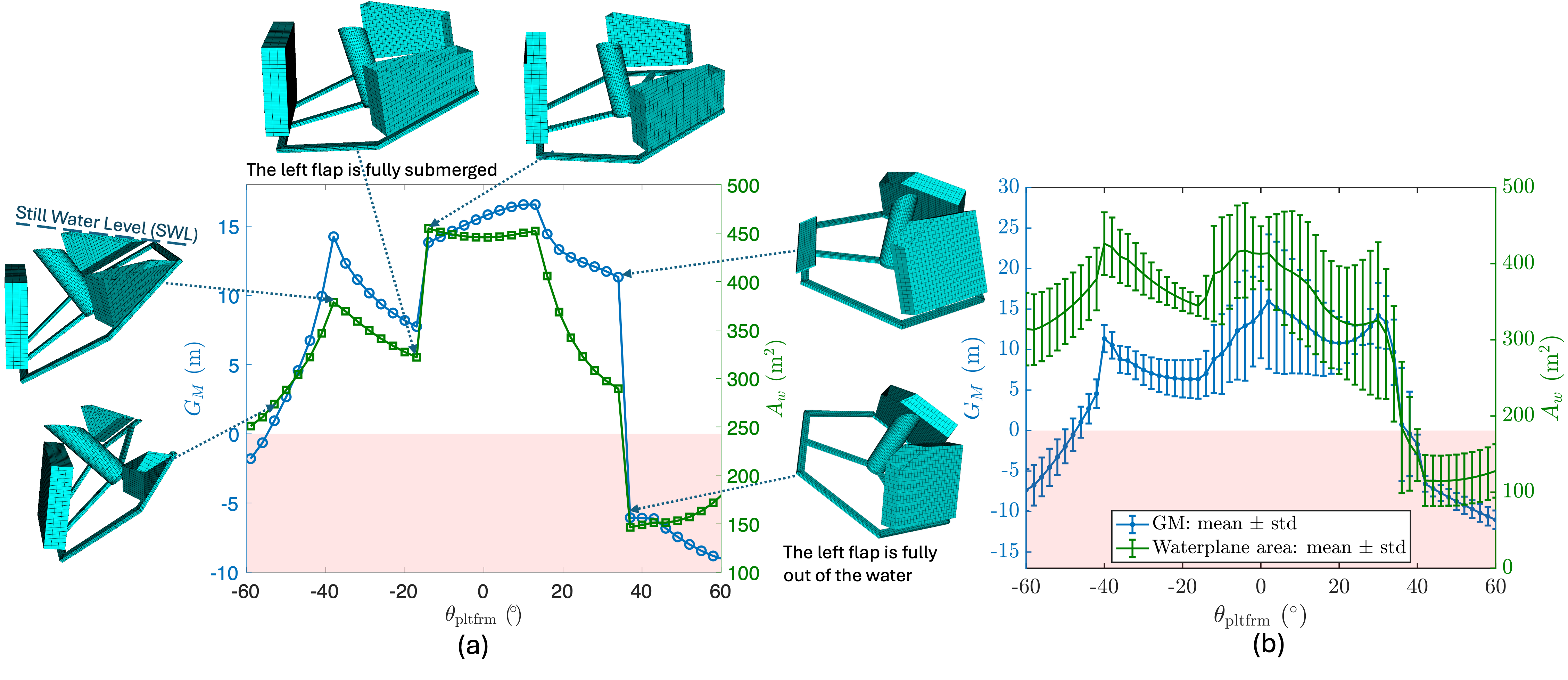}
    %\caption{\small Hydrostatic stability of the hybrid wind–wave platform evaluated using \texttt{Capytaine}. The plot shows the longitudinal metacentric height \(G_M\) (blue, left y-axis) and the corresponding waterplane area (red, right y-axis) as functions of pitch angle \(\theta\), ranging from \(-60^\circ\) to \(+60^\circ\). The shaded region indicates hydrostatic instability where \(G_M < 0\). Insets illustrate the submerged geometry at selected angles. Notably, the trend of \(G_M\) closely follows that of the waterplane area, suggesting that variations in waterplane area are the dominant factor influencing stability in this configuration. The platform remains hydrostatically stable for pitch angles within approximately \(\pm37^\circ\).}
    \caption{\small Hydrostatic stability of the hybrid wind–wave platform evaluated using \texttt{Capytaine}. 
(a) Baseline case with all three flaps fixed in the $+z$ direction while the platform pitch angle $\theta$ is varied from $-60^\circ$ to $+60^\circ$. The longitudinal metacentric height $G_M$ (blue, left $y$-axis) and waterplane area $A_w$ (green, right $y$-axis) are shown, with shaded regions indicating instability where $G_M < 0$. Insets illustrate representative submerged geometries at selected angles. 
(b) Parametric study where, at each pitch angle, the three flaps are independently rotated through $5^3$ orientations from $-40^\circ$ to $+40^\circ$. Plotted are the mean values of $G_M$ and $A_w$ across flap orientations, with shaded bands denoting $\pm$ one standard deviation. Results highlight the strong influence of flap rotation on both $A_w$ and $G_M$, and show that stability is maintained within approximately $\pm 37^\circ$, depending on flap configuration.}
    \label{fig:stability_GM}
\end{figure}

In a separate analysis based on the baseline design, all three flaps were held fixed in the \(+z\) direction, while the platform was rotated through angles ranging from \(-60^{\circ}\) to \(+60^{\circ}\).
Figure~\ref{fig:stability_GM}(a) shows the results of this study, where the $x$-axis represents the 
platform rotation angle (in degrees), the left $y$-axis indicates the metacentric height (in meters), 
and the right $y$-axis shows the waterplane area (in m$^2$). Representative submerged portions of 
the system are also illustrated. As seen, the system becomes unstable below approximately 
$-55^{\circ}$ and above $40^{\circ}$. In addition, a sharp jump in the metacentric height is observed, 
which closely follows the trend of the waterplane area. For example, around $40^{\circ}$ the left flap 
emerges completely from the water, causing a significant reduction in the waterplane area. Similarly, 
near $-20^{\circ}$ the flap becomes fully submerged, again decreasing the waterplane area. 

In another study, the platform pitch angle was varied from $-60^{\circ}$ to $+60^{\circ}$ in increments of $2^{\circ}$ (61 cases). For each platform orientation, the three flaps were independently rotated from $-40^{\circ}$ to $+40^{\circ}$ in increments of $20^{\circ}$ (5 levels each), resulting in a total of $61 \times 5^3$ configurations. For every configuration, hydrostatic properties---including center of mass, buoyancy, and waterplane area---were recalculated, and the longitudinal metacentric height $G_M$ was evaluated. Figure~\ref{fig:stability_GM}(b) summarizes the results. The $x$-axis shows platform pitch angle, the left $y$-axis shows $G_M$ (m), and the right $y$-axis shows the waterplane area $A_w$ (m$^2$). At each platform angle, the mean and standard deviation of $G_M$ and $A_w$ across all flap orientations are plotted. These results show that flap rotation strongly influences both the waterplane area and the metacentric height. By examining the mean and variation, one can identify the pitch angle ranges over which the platform remains stable under different flap configurations. This analysis highlights the dominant role of waterplane area in determining $G_M$ and provides insight into the stability envelope of the hybrid platform. In practice, the waterplane area can be modified by changing flap width, thickness, and related geometric parameters, which directly influence the extent of the waterplane surface.

\subsection{Sensitivity Analysis of Geometric Parameters}
\label{subsec:DOE}

In this section, a sensitivity analysis is conducted to investigate how variations in the geometric parameters of the hybrid system affect hydrostatic stability and key time-domain performance metrics, including power output, platform motion, and tower stress. The analysis involves a parametric sweep in which 12 geometric design variables are each varied independently by $\pm20\%$ around their baseline value, sampled at 9 uniformly spaced levels, while all other variables are held at their baseline values. Table~\ref{tab:design-levels} summarizes the design variables and the corresponding levels used in the study. In total, this results in $12 \times 9 = 108$ unique design cases. While this approach does not explore the full factorial space of $9^{12}$ combinations, it provides a clear assessment of the sensitivity of system responses to individual geometric parameters relative to the chosen baseline design.

\begin{table}[ht!]
  \centering
  \caption{Design variable levels used in the parametric sweep for the sensitivity analysis. Each of the 12 geometric parameters is varied independently from $-20\%$ to $+20\%$ of its baseline value, sampled at nine uniformly spaced levels. The table shows the specific values used for each design variable at each level.}
  \label{tab:design-levels}
  \scriptsize
  \setlength{\tabcolsep}{4pt}
  \begin{tabular}{c
                  rrrrrrrrrrrrr}
    \toprule
    \textbf{Level} &
    $z_{\mathrm{dr}}^p$ &
    $z_{\mathrm{fr}}^p$ &
    $l_s^p$ &
    $w_{xy}^p$ &
    $w_z^p$ &
    $d_c^p$ &
    $l^f$ &
    $h^f$ &
    $w^f$ &
    $l_t$ &
    $d_b^t$ &
    $d_t^t$ \\
    \midrule
     1 & -16.0 &  8.0 & 22.984 & 1.60 & 0.80 & 4.400 & 20.0 & 17.6 & 0.32 &  62.08 & 0.022 & 0.015 \\
     2 & -17.0 &  8.5 & 24.420 & 1.70 & 0.85 & 4.675 & 21.3 & 18.7 & 0.34 &  65.96 & 0.023 & 0.016 \\
     3 & -18.0 &  9.0 & 25.857 & 1.80 & 0.90 & 4.950 & 22.5 & 19.8 & 0.36 &  69.84 & 0.024 & 0.017 \\
     4 & -19.0 &  9.5 & 27.294 & 1.90 & 0.95 & 5.225 & 23.8 & 20.9 & 0.38 &  73.72 & 0.026 & 0.018 \\
     5 & -20.0 & 10.0 & 28.730 & 2.00 & 1.00 & 5.500 & 25.0 & 22.0 & 0.40 &  77.60 & 0.027 & 0.019 \\
     6 & -21.0 & 10.5 & 30.166 & 2.10 & 1.05 & 5.775 & 26.3 & 23.1 & 0.42 &  81.48 & 0.028 & 0.020 \\
     7 & -22.0 & 11.0 & 31.603 & 2.20 & 1.10 & 6.050 & 27.5 & 24.2 & 0.44 &  85.36 & 0.030 & 0.021 \\
     8 & -23.0 & 11.5 & 33.040 & 2.30 & 1.15 & 6.325 & 28.8 & 25.3 & 0.46 &  89.24 & 0.031 & 0.022 \\
     9 & -24.0 & 12.0 & 34.476 & 2.40 & 1.20 & 6.600 & 30.0 & 26.4 & 0.48 &  93.12 & 0.032 & 0.023 \\
    \bottomrule
  \end{tabular}
\end{table}

\begin{figure}[ht!]
    \centering
    \includegraphics[width=0.49\linewidth]{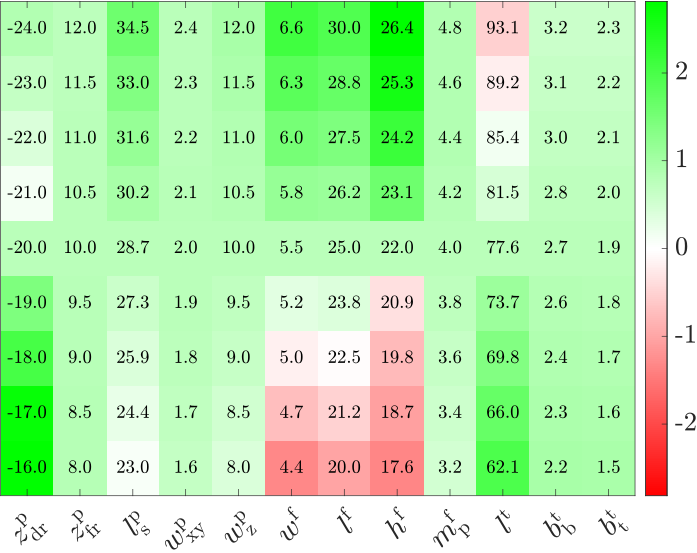}
    \caption{Minimum meta‐centric height obtained over the rotation sweep for each design variable. For each variable (varied from its minimum to maximum value), the meta‐centric height is evaluated at multiple flap rotation angles, and the smallest value of meta-center heights observed is plotted. Negative values indicate an unstable configuration, while positive values indicate a stable one. Note that for the heatmap annotations, the parameter levels for $w_{\mathrm{z}}^{\mathrm{p}}$ (column 5) and $m_{\mathrm{p}}^{\mathrm{f}}$ (column 9) have been multiplied by 10, and those for $b_{\mathrm{b}}^{\mathrm{t}}$ (column 11) and $b_{\mathrm{t}}^{\mathrm{t}}$ (column 12) have been multiplied by 100 to improve readability.
}
    \label{fig:mt_height}
\end{figure}

For the metacentric height sensitivity analysis, each design configuration was evaluated by randomly rotating all three flaps within a range of $\pm 45^{\circ}$, generating 1000 random orientations per case. For each configuration, the minimum metacentric height—that is, the lowest (worst-case) value observed across the 1000 samples—was recorded. Figure~\ref{fig:mt_height} presents the results for all 108 design cases. The figure is organized into 12 columns, each corresponding to one of the geometric design variables, and 9 rows representing the different levels from $-20\%$ to $+20\%$ variation. In each cell, the selected design variable is varied while all others are held at their baseline values (visible in the fifth row). For visualization clarity, the values of $w_{xy}^{\mathrm{p}}$ and $m_{\mathrm{p}}^{\mathrm{f}}$ are multiplied by 10, and those of $b_b^{\mathrm{t}}$ and $b_t^{\mathrm{t}}$ are multiplied by 100. As an example, the cell in the first column and bottom row corresponds to a platform draft of $z_{\mathrm{dr}}^{\mathrm{p}} = -16.0$~m. The color scale indicates the minimum metacentric height obtained under the randomized flap rotations. For this particular configuration, the minimum value is approximately 2~m.

As shown in Fig.~\ref{fig:mt_height}, increasing the flap geometry variables---including flap 
width ($w^{\mathrm{f}}$), length ($l^{\mathrm{f}}$), and height ($h^{\mathrm{f}}$)---has a positive effect on the metacentric height. 
When these parameters increase, the metacentric height tends to larger positive values, whereas 
reductions in these variables lead to smaller or even negative metacentric heights. The underlying 
reason is that increasing flap dimensions enlarges the waterplane area, thereby improving stability.  By contrast, the tower length ($l_t$) has a negative effect: as the tower becomes taller, the total 
system center of gravity rises, reducing the metacentric height. The hexagon side length 
($l_s^{\mathrm{p}}$) also shows a positive effect, again due to its contribution to increasing the waterplane area. 
The influence of the draft ($z_{\mathrm{dr}}^{\mathrm{p}}$) is more complex: the metacentric height first decreases and 
then increases as draft increases. 

Overall, Fig.~\ref{fig:mt_height} provides insight into the order and relative importance of design 
variables for ensuring hydrostatic stability under large flap rotations. For example, small flap 
dimensions yield an unstable system. However, it should be noted that the most negative values 
occur primarily at extreme flap rotations (around $45^{\circ}$). If the allowable flap rotation is 
reduced (e.g., by mechanical stops), even small flap geometries can yield hydrostatically stable designs. From this figure, we can interpret the influence of each design variable on system stability. For instance, increasing the platform draft, flap height, flap length, and flap width generally improves metacentric height and thus hydrostatic stability. In contrast, increasing tower height reduce stability by raising the center of gravity and reducing the waterplane area.

\begin{figure}[ht!]
  \centering
  % first subfigure
  \begin{subfigure}[b]{0.49\linewidth}
    \centering
    \includegraphics[width=\linewidth]{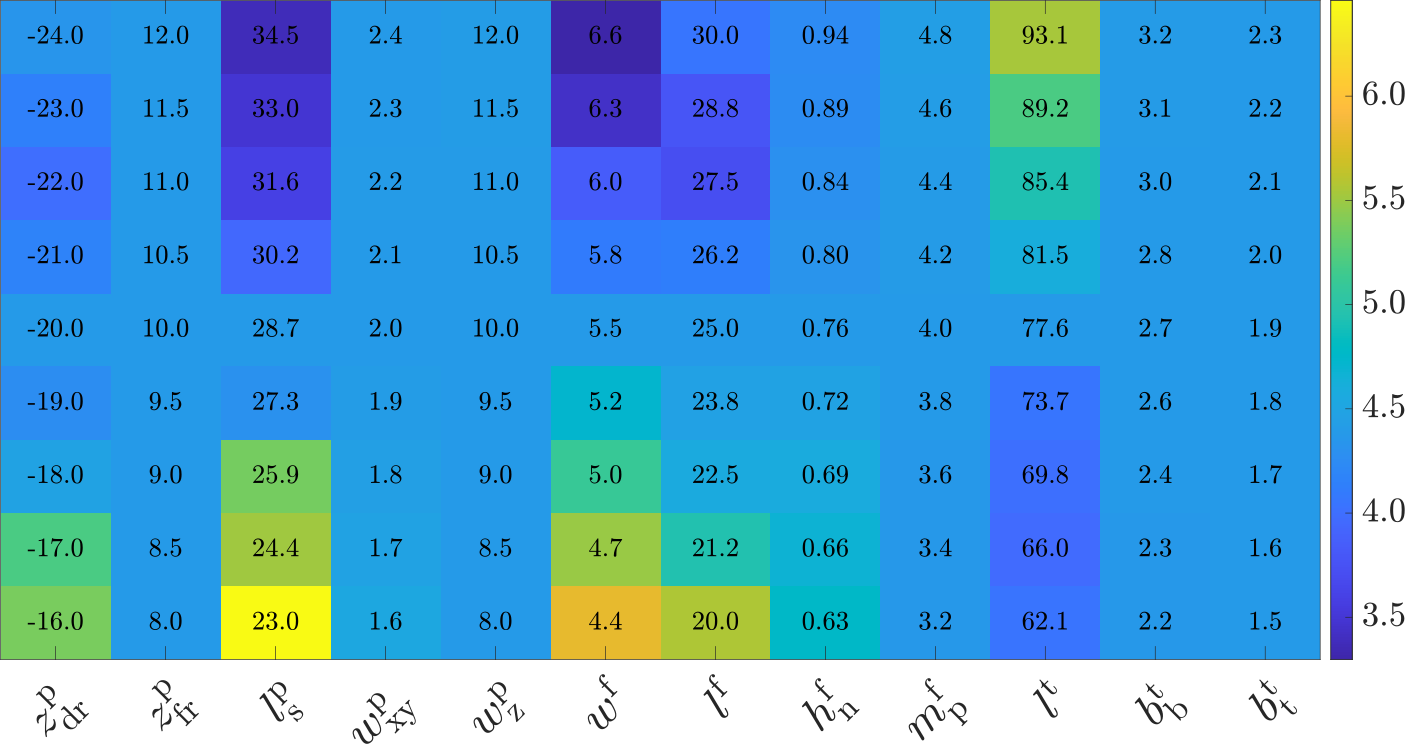}
    \caption{Max platform pitch \([\mathrm{deg}]\).}
    \label{fig:sub1}
  \end{subfigure}
  \hfill
  % second subfigure
  \begin{subfigure}[b]{0.49\linewidth}
    \centering
    \includegraphics[width=\linewidth]{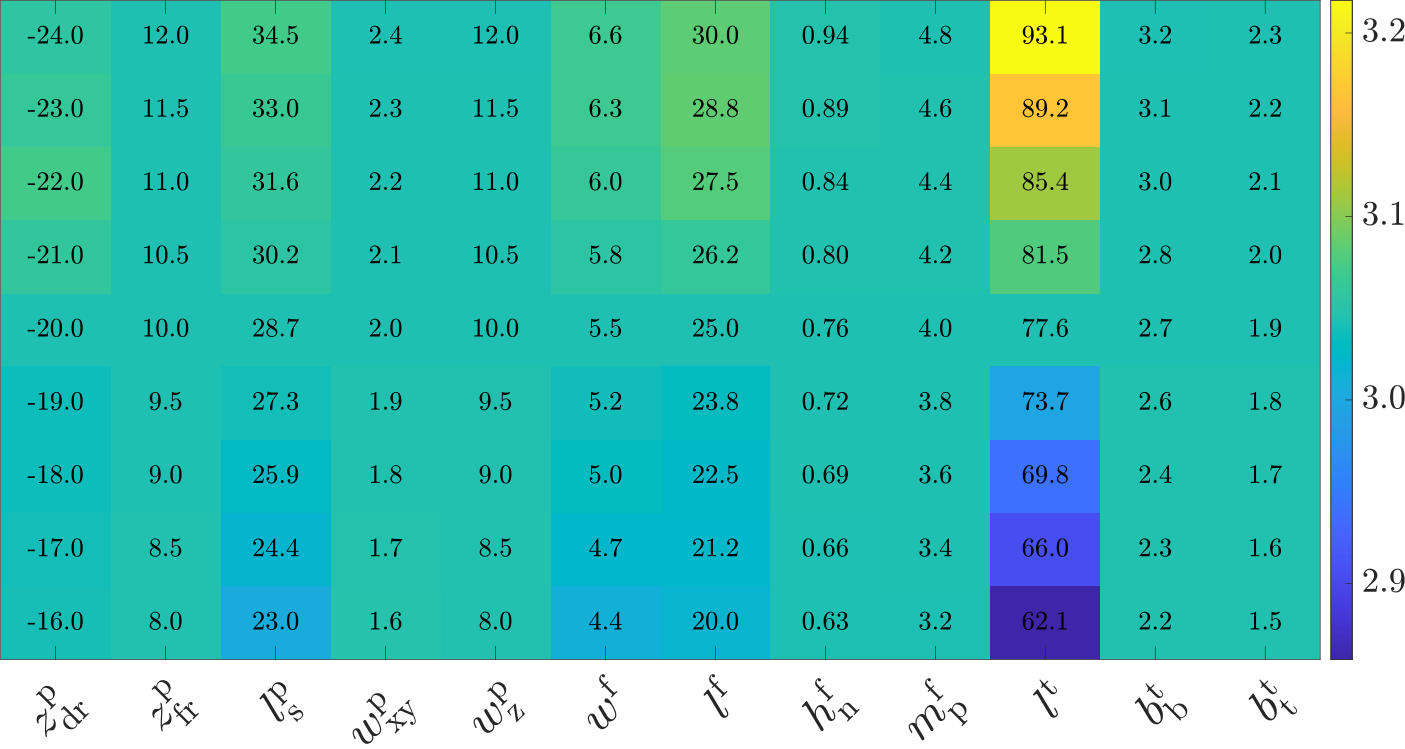}
    \caption{Mean wind turbine power \([\mathrm{MW}]\)}
    \label{fig:sub2}
  \end{subfigure}
  \\
  \begin{subfigure}[b]{0.49\linewidth}
    \centering
    \includegraphics[width=\linewidth]{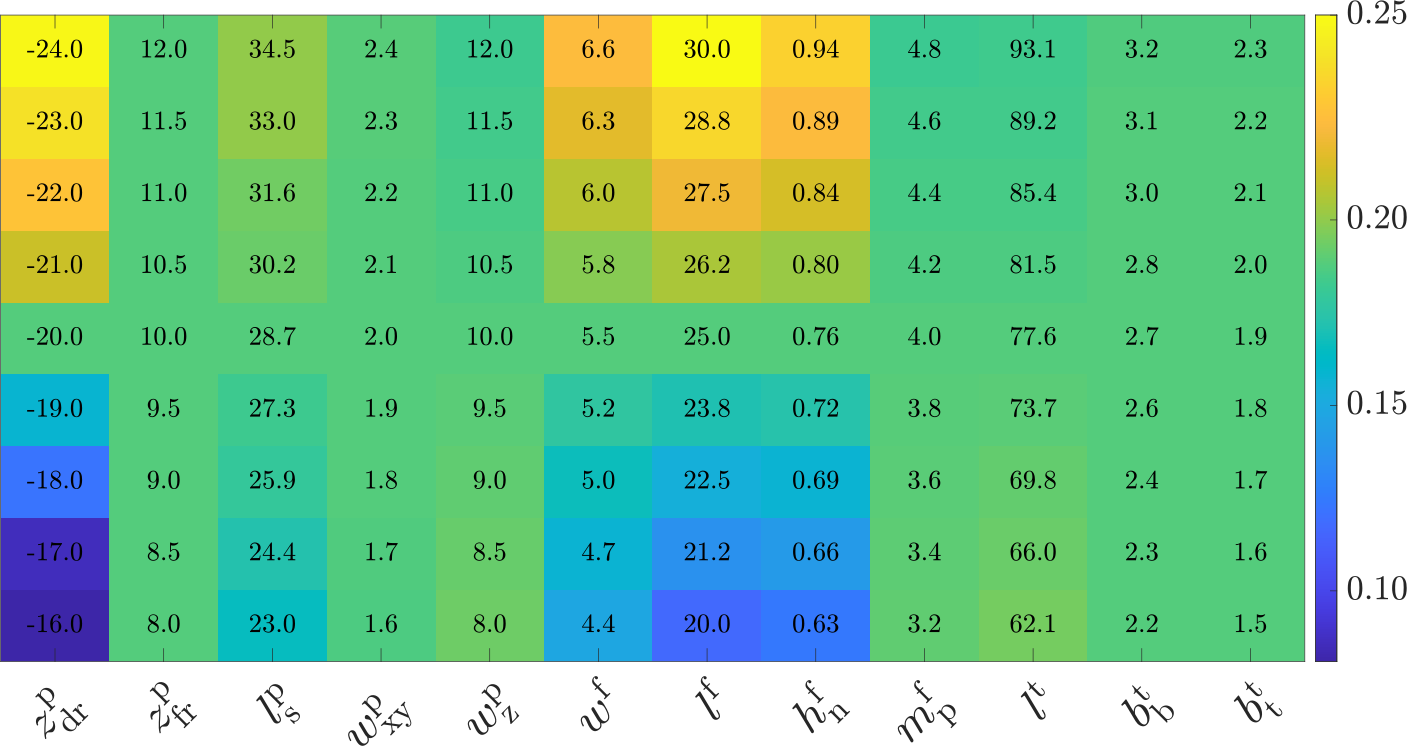}
    \caption{Mean WEC power \([\mathrm{MW}]\)}
    \label{fig:sub1}
  \end{subfigure}
  \hfill
  % second subfigure
  \begin{subfigure}[b]{0.49\linewidth}
    \centering
    \includegraphics[width=\linewidth]{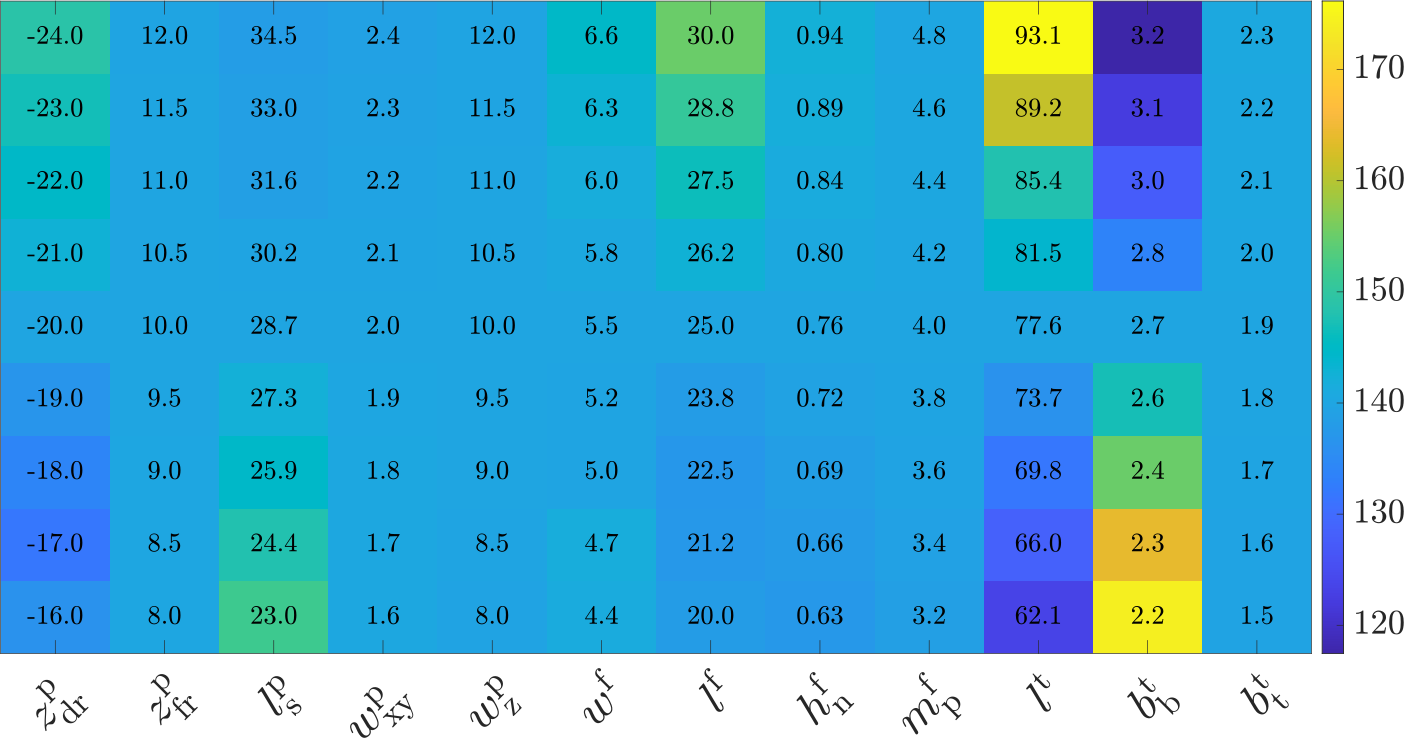}
    \caption{Max bottom stress \(\sigma_{\mathrm{bot}}^{\mathrm{mean}}\) \([\mathrm{MPa}]\).}
    \label{fig:sub2}
  \end{subfigure}
  \caption{Time-domain sensitivity analysis results across 108 design configurations. Each subfigure shows the response of a key performance metric to geometric parameter variation. (a) Maximum platform pitch angle, indicating rotational stability. (b) Mean wind turbine power output under 11.35~m/s wind speed, influenced primarily by hub height. (c) Mean WEC power, strongly affected by flap geometry and platform draft. (d) Maximum bottom stress in the tower, with major contributions from tower length and base thickness. The $x$-axis in all plots represents the design variables varied in the sensitivity analysis.}
  \label{fig:4_plots_heatmaps}
\end{figure}

The next sensitivity analysis results involve simulating the system dynamics in the time domain for each of the 108 design configurations and reporting key outputs such as the maximum platform pitch angle and the average wind turbine power. For the time-domain dynamic simulations, WEC-Sim~\cite{wecsim} was utilized. The $x$-axis in Figure~\ref{fig:4_plots_heatmaps} shows the design variables, with the exception that one of the variables, $h^{\mathrm{f}}_n$, represents the ratio of the submerged flap height to the total flap height. In this study, the submerged portion was fixed at 16.6~m, and variations were achieved by changing the freeboard height.

Figure~\ref{fig:4_plots_heatmaps}(a) presents heatmaps of the maximum platform 
pitch angle. As shown in the legend, the maximum platform pitch across all 108 designs ranges 
from 3$^{\circ}$ to 6.5$^{\circ}$. The dominant effects on platform pitch are associated with the 
hexagon side length ($l_s^{\mathrm{p}}$) and the flap width ($w^{\mathrm{f}}$). Increasing the flap 
height ($h^{\mathrm{f}}$) and flap length ($l^{\mathrm{f}}$), as well as decreasing the tower length 
($l_t$), also have a positive influence in reducing the maximum pitch. The influence of the draft ($z_{\mathrm{dr}}^{\mathrm{p}}$) is more complex and does not follow a consistent trend. The influence of the remaining design variables on the maximum platform 
pitch angle is relatively limited.

Figure~\ref{fig:4_plots_heatmaps}(b) shows the heatmap of the mean wind turbine power. The mean turbine power ranges from approximately 2.8 to 3.2~MW under an incoming wind speed of 11.35~m/s, which lies in Region~2 of the turbine power curve (below the rated wind speed). Among the design variables, the tower length ($l^t$) has the dominant effect: as the tower length increases, the effective hub height increases, exposing the rotor to higher wind speeds and thereby increasing turbine power output. The flap width ($w^{\mathrm{f}}$) and hexagon side length ($l_s^{\mathrm{p}}$) also exert smaller secondary effects, while the influence of the remaining design variables is minimal.

Figure~\ref{fig:4_plots_heatmaps}(c) presents the heatmap of the mean WEC power, calculated as the sum of the time-averaged power from all three flaps divided by three. As shown, the 
mean WEC power ranges from 0.10 to 0.25~MW across all 108 geometries. The largest influence 
on the mean WEC power comes from the flap geometry variables, including flap width 
($w^{\mathrm{f}}$), length ($l^{\mathrm{f}}$), and height ($h^{\mathrm{f}}$), as well as the draft 
($z_{\mathrm{dr}}^{\mathrm{p}}$). Increasing the flap dimensions or the draft results in higher 
WEC power, which is consistent with physical expectations. The hexagon side length 
($l_s^{\mathrm{p}}$) also has a smaller but noticeable effect, while the influence of the remaining 
design variables is very limited.

Figure~\ref{fig:4_plots_heatmaps}(d) shows the heatmap of the tower max bottom stress. As shown, 
the max bottom stress varies from 120 to 170~MPa across all 108 design configurations. The 
tower length ($l_t$) and tower base thickness ($b_b^{\mathrm{t}}$) have the most significant effects. 
As the tower length increases, the effective hub height increases, resulting in higher incident 
wind speeds; this, in turn, increases the thrust force and consequently the tower stress. In contrast, 
increasing the tower base thickness enlarges the cross-sectional area, thereby reducing the mean 
bottom stress. The flap geometry and the hexagon side length ($l_s^{\mathrm{p}}$) also have smaller 
secondary effects, while the influence of the other design variables is very limited.

Based on the sensitivity analysis results, an improved design can be recommended by enlarging the flap geometry (width, length, and height) and slightly increasing the platform draft, as these changes consistently improved hydrostatic stability and wave power capture. At the same time, tower height presents a clear trade-off: while a taller tower is beneficial for wind energy extraction, it has a destabilizing effect due to raising the system’s center of gravity and increasing structural loading. To balance these effects, a reduced tower height combined with a thicker tower base can mitigate structural stresses while still maintaining sufficient hub height for wind power production. A representative trade-off design would therefore feature larger flaps with higher submerged portions, a moderately deeper draft, and a shorter but stiffer tower. While the remainder of this paper continues to use the baseline configuration for comparison with the NREL semi-submersible reference, these sensitivity-driven trends suggest a clear pathway toward more stable and better-performing hybrid designs. Future work will refine this recommended configuration through integrated optimization and additional simulations under irregular sea states.

%In summary, the DOE analysis reveals that flap geometry (width, length, and height), platform draft, and hexagon side length are the most influential parameters for improving hydrostatic stability and maximizing wave energy capture. Tower height, while beneficial for wind energy extraction, has a destabilizing effect due to its influence on the system's center of gravity and structural loading. These trends highlight key design trade-offs between platform stability, energy performance, and structural response, providing valuable guidance for optimizing future hybrid wind--wave energy platforms.

\subsection{Time-Domain Response Under Two Wave Directions}
\label{subsec:Time}

This section examines the time-domain response of the hybrid system using the baseline design under two different wave directions: $0^\circ$ and $60^\circ$. In both cases, the incoming wind direction is fixed at $0^\circ$, so only the wave direction is varied. For the $0^\circ$ case, waves propagate along the surge direction. As a result, the left flap (Flap~1)—which is oriented perpendicular to the wave front—extracts the most power, while Flaps~2 and~3 extract comparatively less. In the $60^\circ$ case, the wave direction aligns with Flap~3, enabling it to extract the maximum power, while the left and bottom flaps generate lower output.

\begin{figure}[ht!]
  \centering
  % first subfigure
  \begin{subfigure}[b]{0.4\linewidth}
    \centering
    \includegraphics[width=\linewidth]{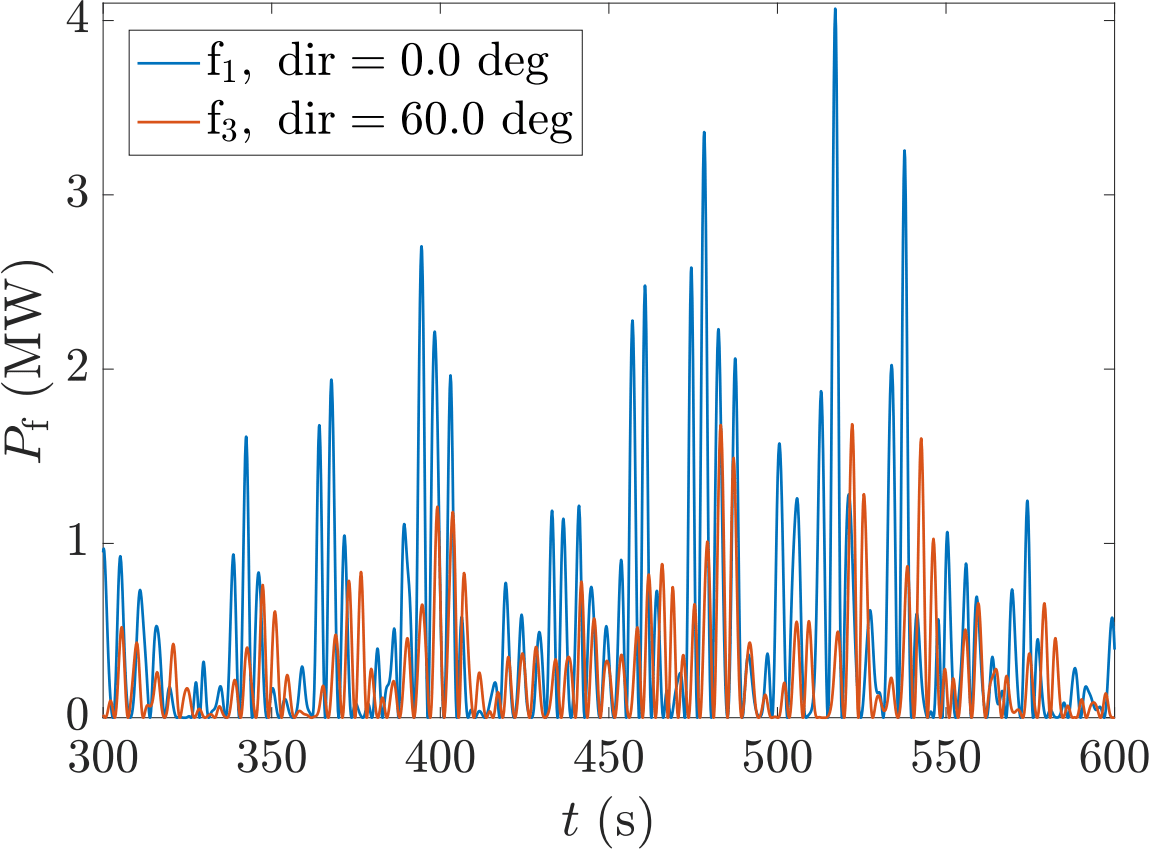}
    \caption{}
    \label{fig:sub1}
  \end{subfigure}
  %\hfill
  % second subfigure
  \begin{subfigure}[b]{0.4\linewidth}
    \centering
    \includegraphics[width=\linewidth]{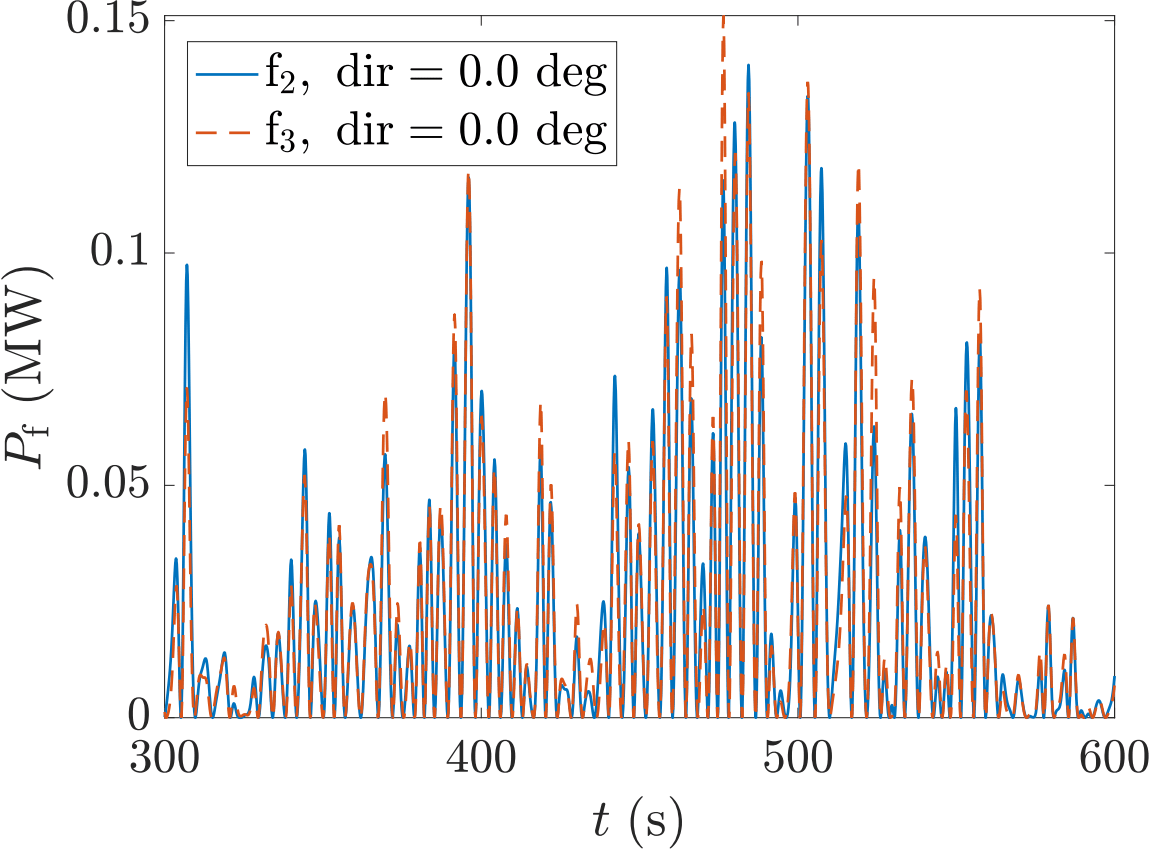}
    \caption{}
    \label{fig:sub2}
  \end{subfigure}
  \\
  \begin{subfigure}[b]{0.4\linewidth}
    \centering
    \includegraphics[width=\linewidth]{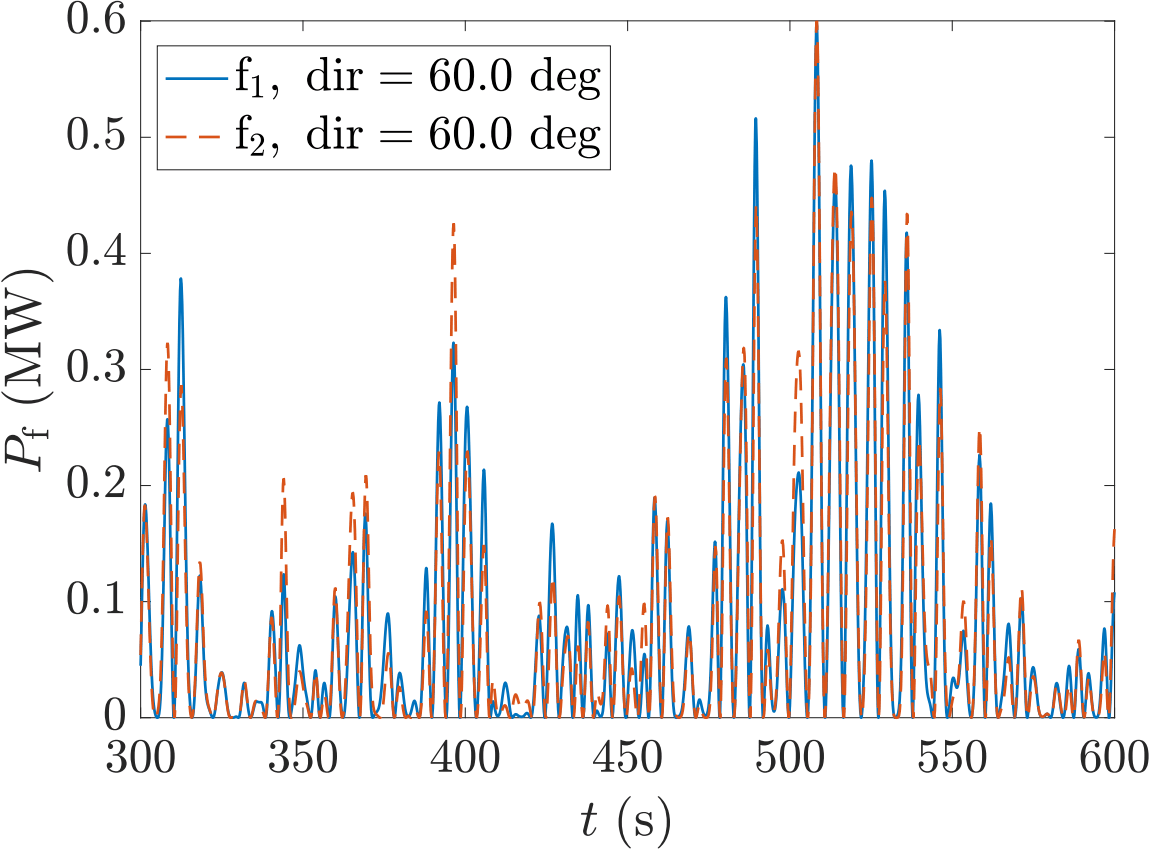}
    \caption{}
    \label{fig:sub1}
  \end{subfigure}
  %\hfill
  % second subfigure
  \begin{subfigure}[b]{0.4\linewidth}
    \centering
    \includegraphics[width=\linewidth]{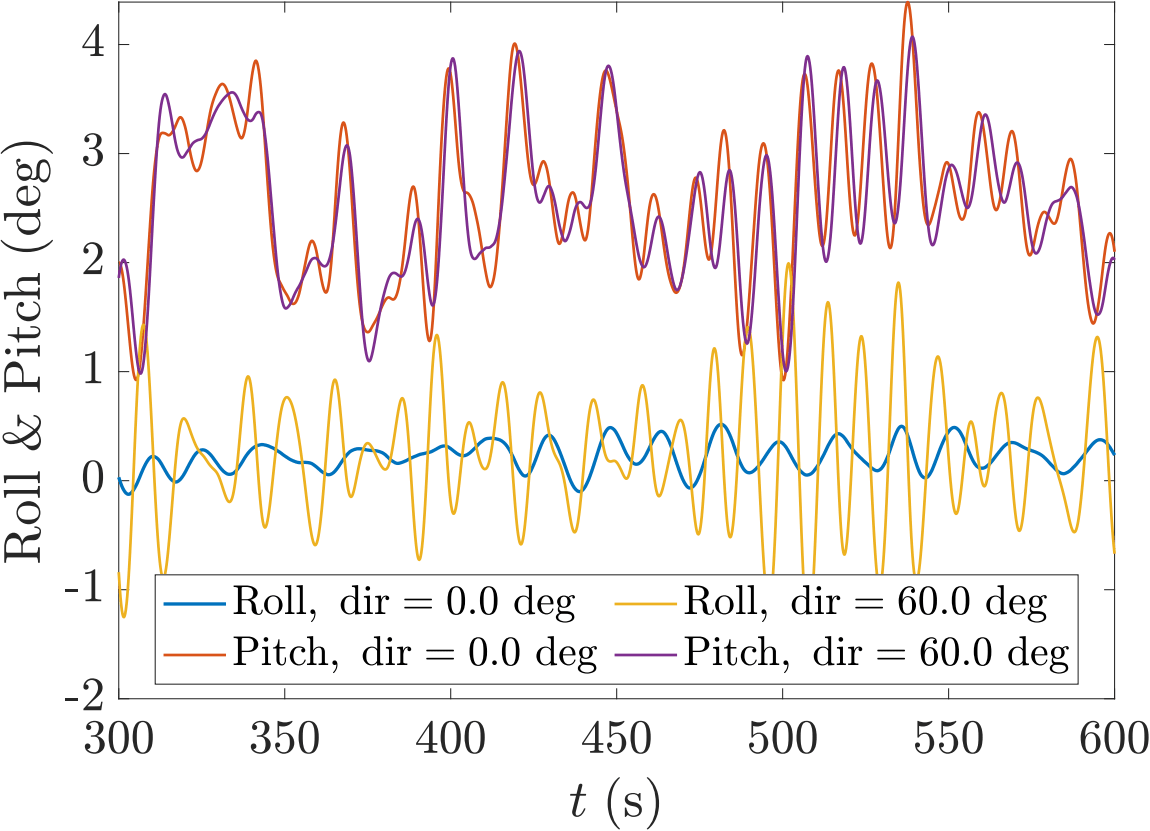}
    \caption{}
    \label{fig:sub2}
  \end{subfigure}
\caption{Time-domain simulation results of the hybrid platform under two wave incidence angles: 0° and 60°. 
(a) Power output of the flap aligned perpendicular to the wave direction in each case—flap 1 for 0° waves and flap 3 for 60° waves. 
(b) Power output of flaps 2 and 3 under 0° wave incidence, which are symmetric about the wave direction and show nearly identical performance. 
(c) Power output of flaps 1 and 2 under 60° waves, also exhibiting similar behavior and significantly lower power compared to flap 3. 
(d) Platform pitch and roll motions for both cases. The 0° case results in minimal roll (about 0.5°) and moderate pitch (up to 4.4°), while the 60° case shows increased roll (up to 1.8°) and slightly lower pitch (about 4.0°).}
  \label{fig:4_plots_deg}
\end{figure}

Figure~\ref{fig:4_plots_deg}(a) shows the power output of the flap that performs best in each case: the flap 1 under 0° waves and the flap 3 under 60° waves. Figure~\ref{fig:4_plots_deg}(b) displays the power outputs of the flaps 2 and 3 under 0° waves. These results are nearly identical due to the system's geometric symmetry with respect to the wave direction. Similarly, Figure~\ref{fig:4_plots_deg}(c) presents the power output of the flaps 1 and 2 under 60° waves, again demonstrating comparable performance and significantly lower power compared to the top flap. Figure~\ref{fig:4_plots_deg}(d) shows the platform's roll and pitch motions for both wave directions. In the 0° case, the roll remains minimal (approximately 0.5°) since the wave propagates along the \(x\)-axis, preserving transverse symmetry. In contrast, the 60° case introduces asymmetry in the wave loading, resulting in increased roll motion (up to 1.8°). The corresponding pitch motions are 4.4° and 4.0° for the 0° and 60° cases, respectively, indicating slightly reduced pitching under oblique wave incidence.

Table~\ref{tab:deg_study} presents the average power extracted by the 3 flaps under two wave incidence scenarios. In the first scenario (0° wave angle), the  flap 1 generates an average of 0.47~MW, while the flaps 2 and 3 each extract about 0.02~MW. In the second scenario (60° wave angle), the flap 3 produces an average of 0.24~MW, with the left and bottom flaps generating approximately 0.08~MW each. The total WEC power is 0.51~MW for the 0° case and 0.40~MW for the 60° case. The corresponding average wind turbine power outputs are 3.26~MW and 3.25~MW, respectively, under a mean hub-height wind speed of 11.35~m/s. The table also includes the average and maximum von Mises stress at the tower base, as well as the peak pitch and roll angles for each scenario.

\begin{table}[ht!]
  \centering
 \caption{Summary of performance metrics for two wave incidence directions (0° and 60°). The table reports: average power extracted by each flap ($P_f^1$, $P_f^2$, $P_f^3$), total WEC power ($\sum P_f$), average wind turbine power ($P_t$), mean and maximum von Mises stress at the tower base ($\sigma_{\mathrm{bot}}^{\mathrm{mean}}$, $\sigma_{\mathrm{bot}}^{\mathrm{max}}$), and maximum pitch and roll angles of the platform ($\mathrm{Pitch}^{\max}$, $\mathrm{Roll}^{\max}$).}
  \scalebox{0.75}{ % or \footnotesize if you need it smaller
  \begin{tabular}{lccccccccc}
    \toprule
    Condition
      & $P_f^1\,[\mathrm{MW}]$
      & $P_f^2\,[\mathrm{MW}]$
      & $P_f^3\,[\mathrm{MW}]$
      & $\sum P_f\,[\mathrm{MW}]$
      & $P_t\,[\mathrm{MW}]$
      & $\sigma_{\mathrm{bot}}^{\mathrm{mean}}\,[\mathrm{MPa}]$
      & $\sigma_{\mathrm{bot}}^{\mathrm{max}}\,[\mathrm{MPa}]$
      & $\mathrm{Pitch}^{\max}\,[\deg]$
      & $\mathrm{Roll}^{\max}\,[\deg]$ \\
    \midrule
    $\deg=0$  
      & 0.47 & 0.02 & 0.02 & 0.51 & 3.26 & 75 & 140 & 4.4 & 0.5 \\
    $\deg=60$ 
      & 0.08 & 0.08 & 0.24 & 0.40 & 3.25 & 77 & 141 & 4.0 & 1.8 \\
    \bottomrule
  \end{tabular}}
  \label{tab:deg_study}
\end{table}

\begin{figure}[ht!]
  \centering
  % first subfigure
  \begin{subfigure}[b]{0.49\linewidth}
    \centering
    \includegraphics[width=\linewidth]{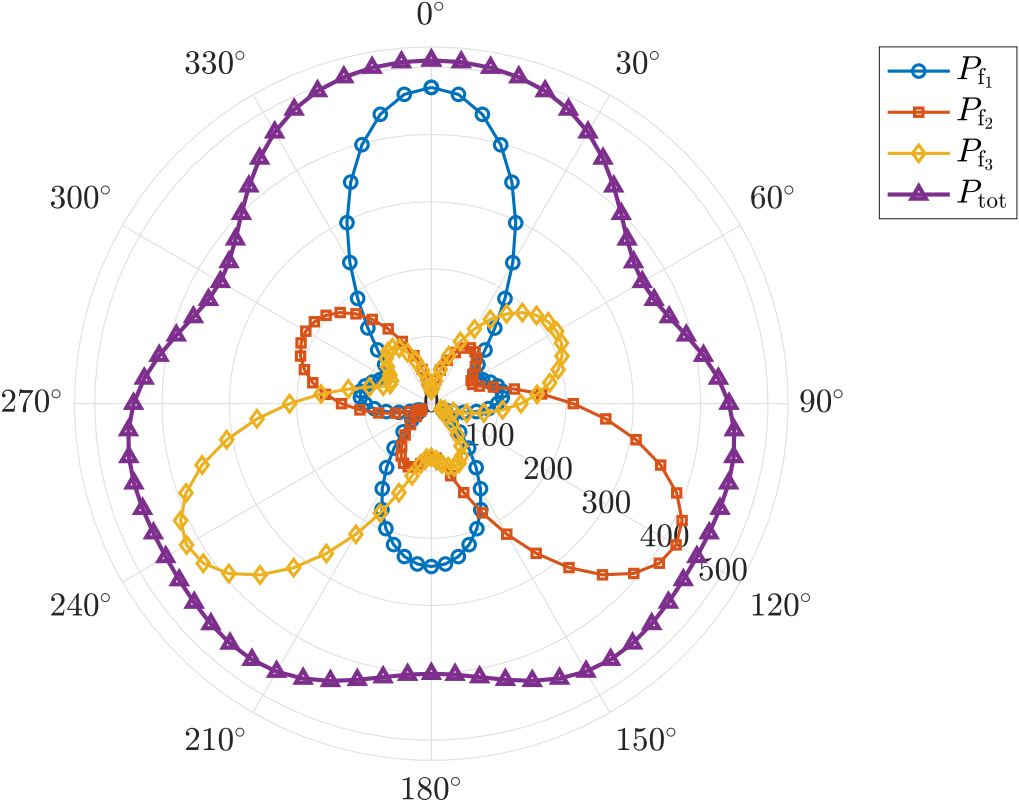}
    \caption{}
  \end{subfigure}
  %\hfill
  % second subfigure
  \begin{subfigure}[b]{0.40\linewidth}
    \centering
    \includegraphics[width=\linewidth]{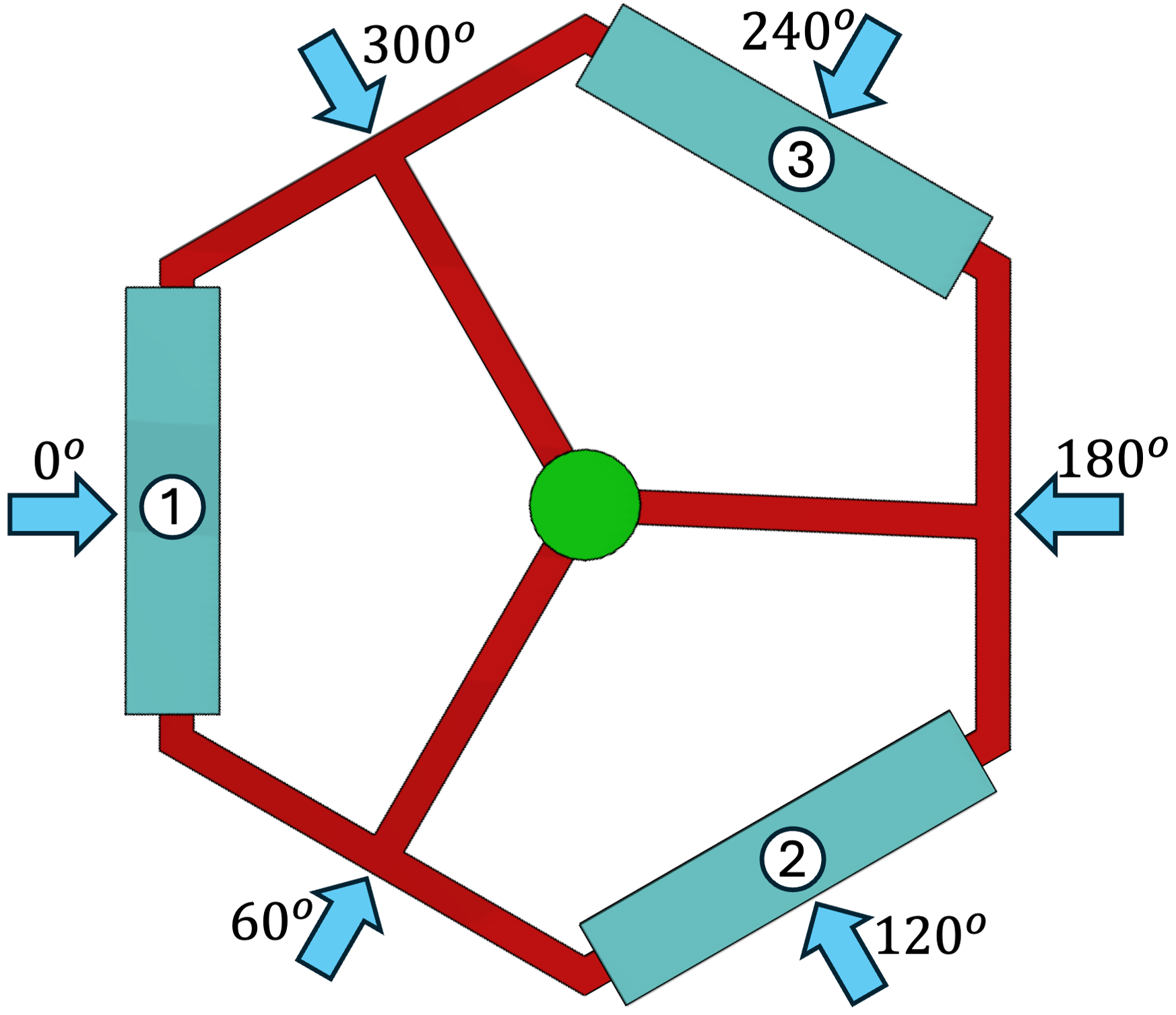}
    \caption{}
  \end{subfigure}
  \caption{(a) Polar plot of the mean power absorbed by each flap ($P_{f_i}$) and the total power ($P_{\mathrm{tot}}$) of the WECs as a function of wave heading angle. Each flap reaches its maximum power when the incident waves are perpendicular to its surface ($0^\circ$, $120^\circ$, and $240^\circ$ for flaps~1–3, respectively), while the corresponding minima occur when another flap is maximally excited. (b) Definition of the wave heading angles, where $0^\circ$ corresponds to waves aligned with the surge direction and angles increase counterclockwise.}
  \label{fig:direction_stuy}
\end{figure}

To further investigate the influence of wave direction on flap power output, we varied the incident wave angle from $0^\circ$ to $360^\circ$ in increments of $5^\circ$. For each case, WEC-Sim simulations were conducted, and the mean power absorbed by each flap as well as the total mean power across all three flaps were recorded. The results are shown in Fig.~\ref{fig:direction_stuy}(a). In this polar plot, the radial axis represents the absorbed power, the angular axis corresponds to the wave heading, and the legend identifies $P_{f_i}$ as the mean power absorbed by the $i$th flap (kW) and $P_{\mathrm{tot}}$ as the total absorbed power. Figure~\ref{fig:direction_stuy}(b) illustrates the definition of the wave heading angles from $0^\circ$ to $360^\circ$. Here, $0^\circ$ corresponds to waves aligned with the surge direction, and as the heading angle increases counterclockwise, the incident wave direction rotates accordingly.

%\begin{figure}[ht!]
%    \centering
%    \includegraphics[width=0.6\linewidth]{wave_dierec_circular.eps}
%    \caption{Mean power absorbed by each flap (\(P_{f_i}\)) and the total power (\(P_{\mathrm{tot}}\)) of the WECs as a function of wave heading angle.}
%    \label{fig:direction_stuy}
%\end{figure}

As shown in Fig.~\ref{fig:direction_stuy}(b), flaps~1–3 reach their maximum power at wave headings of $0^\circ$, $120^\circ$, and $240^\circ$, respectively. This behavior is expected, since at these angles the incident waves are perpendicular to the flap surfaces, enabling each flap to extract the maximum energy from the waves. The maximum total power also occurs at $0^\circ$, with a value of approximately 510~kW. However, when the waves approach from the opposite direction (i.e., $180^\circ$ apart), the absorbed power decreases. For instance, comparing flap~1 at $0^\circ$ and $180^\circ$, the absorbed power at $180^\circ$ is significantly smaller. This reduction occurs because, at $180^\circ$, the wave energy incident on flap~1 is partially shadowed by the other flaps and the central cylinder. A similar trend is observed for flaps~2 and~3, where reduced power occurs at $300^\circ$ and $60^\circ$, respectively.

In Fig.~\ref{fig:direction_stuy}(a), the minimum power for flap~1 occurs at wave headings of $120^\circ$ and $240^\circ$, which correspond to the headings where one of the other two flaps achieves its maximum power. The same observation holds for the minimum power values of flaps~2 and~3. One might expect that the minimum power for flap~1 would occur at $90^\circ$, since the incident waves are aligned with the flap surface. However, the results show that although the power from flap~1 at $90^\circ$ is relatively small (98~kW), it is still larger than the power at $120^\circ$ or $240^\circ$ (19~kW).  Two factors may help explain this behavior. First, when the wave heading is at $120^\circ$ or $240^\circ$, much of the wave energy is absorbed by the corresponding perpendicular flap, leaving less energy available for flap~1. For example, comparing $60^\circ$ and $240^\circ$: in both cases the waves are perpendicular to flap~3, yet flap~3 generates more power at $240^\circ$ than at $60^\circ$. Correspondingly, flap~1 produces only 19~kW at $240^\circ$ compared to 78~kW at $60^\circ$. This indicates that when flap~3 extracts more power from the waves, the power absorbed by flap~1 decreases.  Second, at $90^\circ$ the waves are aligned with flap~1, so one might expect its power to be zero. However, the motions of the platform and other flaps disturb the incident wave field, resulting in nonzero power at this angle. To fully understand the fluid dynamics in the vicinity of the platform, future studies employing high-fidelity computational fluid dynamics tools (e.g., OpenFOAM) are needed. 

These results also highlight the presence of a hydrodynamic shadow effect, where the hexagonal platform, central cylinder, and other structural components reduce the incident wave energy reaching the side flaps. While the present model includes linear hydrodynamic coupling between the platform and flaps, the observed shadowing emphasizes the importance of these interactions. Future work will extend the modeling to include potential nonlinear and viscous effects, as well as flap--platform coupling in greater detail.

\subsection{Feasibility of Platform Pitch Modulation via Flap Angle Sweep}
\label{subsec:flp_cntrl}

One additional benefit of integrating flaps into a floating wind turbine platform—beyond wave energy extraction—is their potential for actively controlling platform pitch motion. This can be examined by prescribing fixed flap angles to modify the submerged volume and thus the buoyancy. In this study, we held Flap~1 at prescribed angles ranging from $-55^\circ$ to $55^\circ$ in increments of $1^\circ$ and measured the steady-state platform pitch response for each case. For this analysis, the wave height was set to zero and the wind speed was fixed at a constant value of $3$~m/s. The PTO proportional and integral gains were set to sufficiently large values to maintain the flap in its prescribed position. Each WEC-Sim simulation was run for $600$~s, and the average value over the final $100$~s of the simulation was reported as the steady-state response. We also recorded the individual torque contributions from Flap~1 due to buoyancy and weight, along with their summation. As the flap angle was varied across the prescribed range, its buoyancy force, the center of buoyancy, and the center of gravity all changed, resulting in a net torque from Flap~1 that directly influenced the platform pitch.

\begin{figure}[ht!]
  \centering
  % first subfigure
  \begin{subfigure}[b]{0.49\linewidth}
    \centering
    \includegraphics[width=\linewidth]{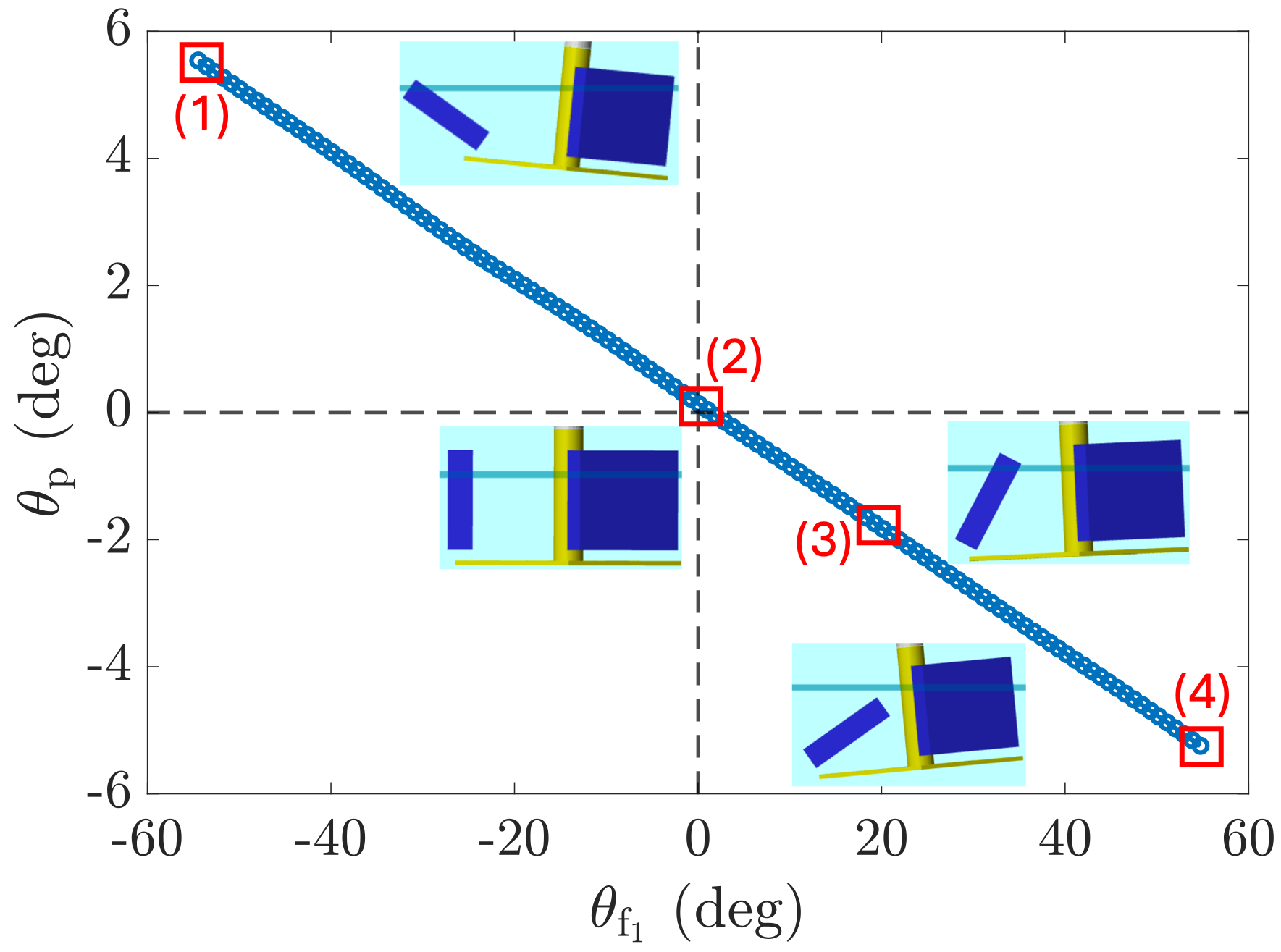}
    \caption{}
    \label{fig:sub1}
  \end{subfigure}
  \hfill
  % second subfigure
  \begin{subfigure}[b]{0.49\linewidth}
    \centering
    \includegraphics[width=\linewidth]{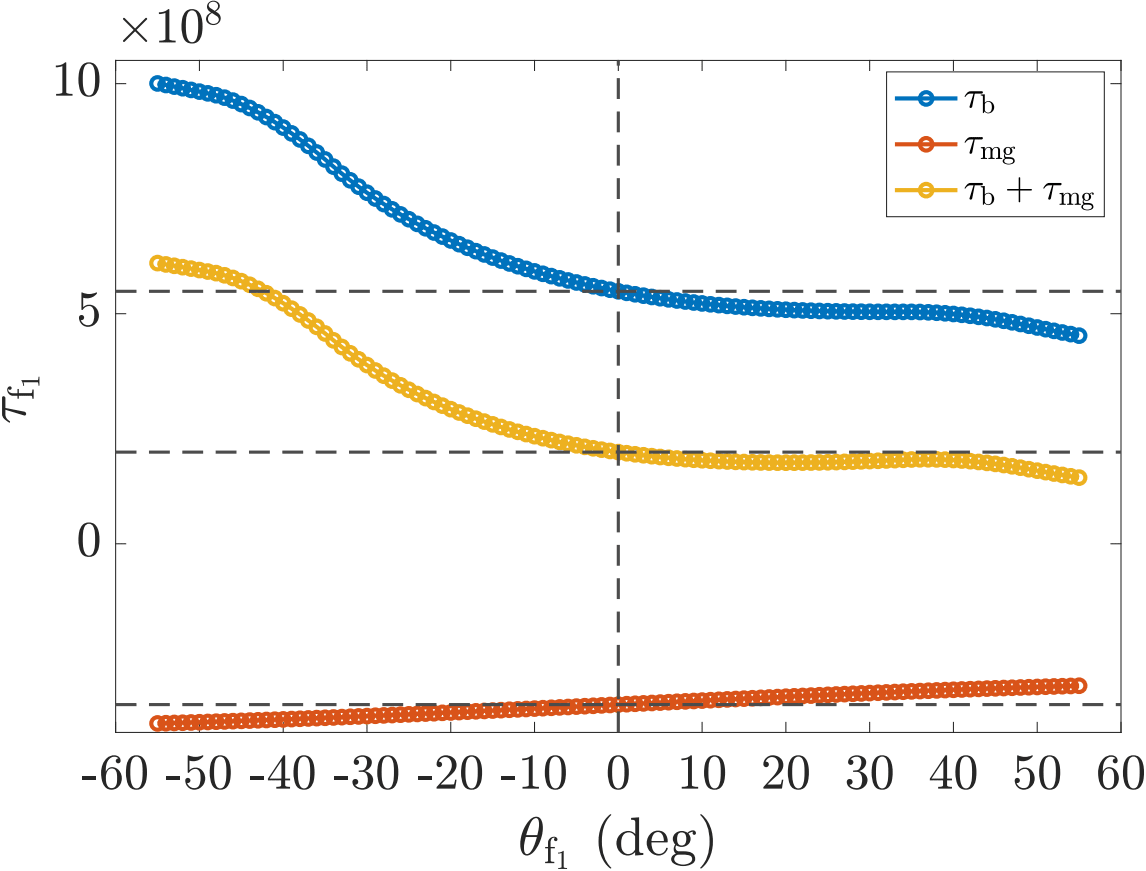}
    \caption{}
    \label{fig:sub2}
  \end{subfigure}
  \caption{Effect of prescribed flap angle on platform pitch. 
(a) Platform pitch angle as a function of Flap~1 angle, with inset panels (1)–(4) showing representative geometries. The pitch shifts from $+5.54^\circ$ at $-55^\circ$ to $-5.24^\circ$ at $+55^\circ$, demonstrating that flap actuation can counteract pitch motion. 
(b) Torque contributions from Flap~1 about the platform COG, including buoyancy torque $\tau_b$, weight torque $\tau_{mg}$, and their summation.}
  \label{fig:4_plots_rot_deg}
\end{figure}

Figure~\ref{fig:4_plots_rot_deg} illustrates this concept. In Fig.~\ref{fig:4_plots_rot_deg}(a), the $x$-axis represents the prescribed flap angle, while the $y$-axis shows the resulting steady-state platform pitch angle. The inset panels labeled (1)–(4) display snapshots of the platform and flap geometry at four sample angles. As the flap angle is swept from $-55^\circ$ to $55^\circ$, the platform pitch angle changes from $5.54^\circ$ to $-5.24^\circ$. Here, negative values correspond to counterclockwise flap orientation, while positive values correspond to clockwise orientation.

In Fig.~\ref{fig:4_plots_rot_deg}(b), the $x$-axis represents the flap angle and the $y$-axis shows the torque from Flap~1 about the platform center of gravity (COG). The legend identifies the buoyancy torque $\tau_b$, the weight torque $\tau_{mg}$, and their summation. When the flap angle is set from $0^\circ$ to $-55^\circ$, $\tau_b$ increases because (1) the submerged volume of the flap increases, which enhances the buoyancy force, and (2) the center of buoyancy (COB) shifts farther from the platform COG, increasing the moment arm. In contrast, when the flap angle is set from $0^\circ$ to $55^\circ$, the buoyancy force still increases, but the distance between the COB and the platform COG decreases, resulting in a reduction of the net buoyancy torque.

The effect of the weight torque of Flap~1 on the platform COG is also shown in Fig.~\ref{fig:4_plots_rot_deg}(b). As the flap angle changes from $-60^\circ$ to $60^\circ$, the distance between the flap COG and the platform COG decreases, and as a result the absolute value of the weight torque $\tau_{mg}$ also decreases. The summation of the buoyancy and weight torques is shown in Fig.~\ref{fig:4_plots_rot_deg}(b) as well. The variation in the combined torque is larger for negative flap angles compared to positive flap angles. Consistent with Fig.~\ref{fig:4_plots_rot_deg}(a), the platform exhibits a slightly higher pitch angle at negative flap angles than at positive flap angles ($5.54^\circ$ vs.~$-5.24^\circ$). It should be noted, however, that these are not the only torques acting on the platform. As the platform rotates, the buoyancy, center of buoyancy, and center of gravity of the other flaps also change, which contributes additional effects on the overall pitch angle.

These results highlight that prescribed flap angles can be used to modulate platform pitch and enhance platform stability. For example, if the platform is rotated by approximately $+5^\circ$ due to wind and large waves, setting Flap~1 to $+55^\circ$ can bring the resulting platform pitch angle close to zero. It should be noted that in this study we only considered Flap~1, but a similar effect can also be achieved by adjusting Flaps~2 and~3.

%\begin{figure}[ht!]
%  \centering
  % first subfigure
%  \begin{subfigure}[b]{0.4\linewidth}
%    \centering
%    \includegraphics[width=0.75\linewidth]{rot_1.png}
%    \caption{}
%    \label{fig:sub1}
%  \end{subfigure}
  %\hfill
  % second subfigure
%  \begin{subfigure}[b]{0.35\linewidth}
%    \centering
%    \includegraphics[width=\linewidth]{rot_2.png}
%    \caption{}
%    \label{fig:sub2}
%  \end{subfigure}
%  \\
%  \begin{subfigure}[b]{0.35\linewidth}
%    \centering
%    \includegraphics[width=\linewidth]{flap_bouyancy_vs_theta.eps}
%    \caption{}
%    \label{fig:sub1}
%  \end{subfigure}
  %\hfill
  % second subfigure
%  \begin{subfigure}[b]{0.35\linewidth}
%    \centering
%    \includegraphics[width=\linewidth]{flap_pltfrm_pitch_study_wwec1.eps}
%    \caption{}
%    \label{fig:sub2}
%  \end{subfigure}
%  \caption{Active control of platform pitch through flap rotation. (a) Rotating the left flap downward increases its buoyancy and induces a clockwise pitch moment. (b) Rotating the top and bottom flaps downward generates an opposing pitch moment. (c) Buoyant force of a flap as a function of rotation angle, showing increased force with greater submersion. (d) Average platform pitch angle from WEC-Sim simulations with flap-1 angles swept from $-40$\textdegree{} to $+46.5$\textdegree{}, demonstrating controllability of pitch via flap actuation.}
%  \label{fig:4_plots_rot_deg}
%\end{figure}

\subsection{Annual Energy Production of Wind and Wave Subsystems}
\label{subsec:wec_wt}

\begin{figure}[ht!]
  \centering
  % first subfigure
  \begin{subfigure}[b]{0.49\linewidth}
    \centering
    \includegraphics[width=\linewidth]{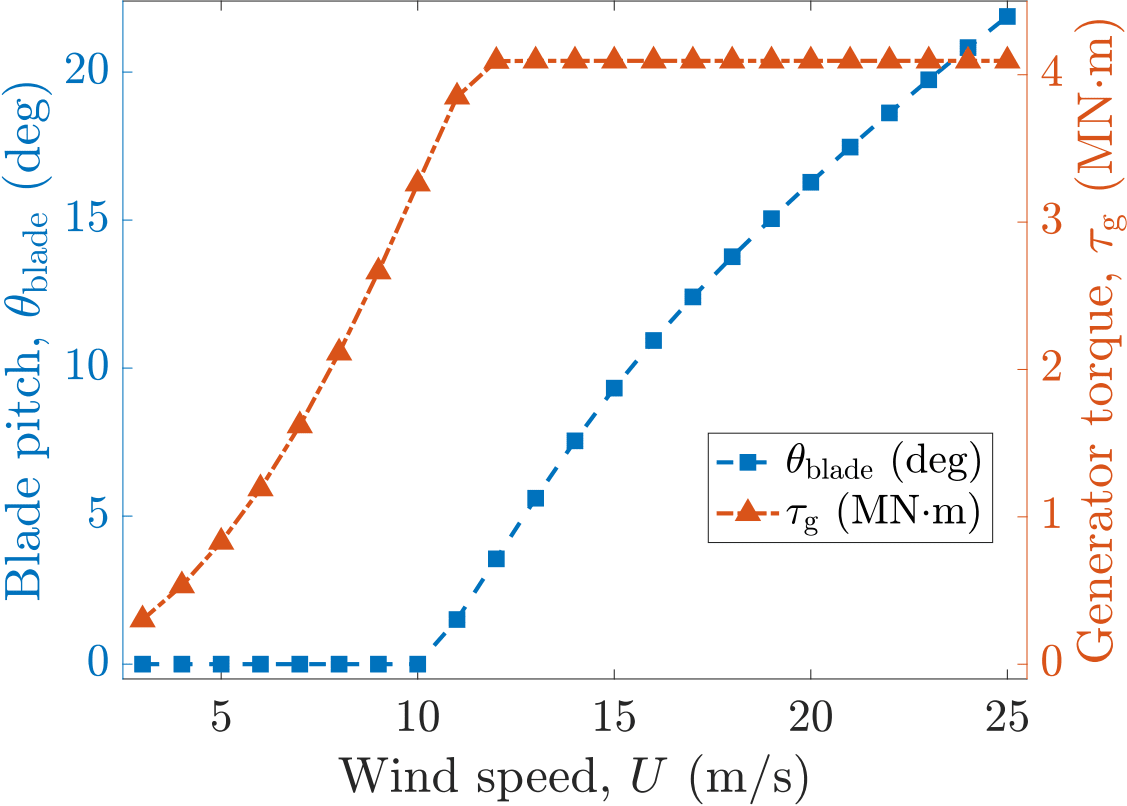}
    \caption{}
    \label{fig:sub1}
  \end{subfigure}
  \hfill
  % second subfigure
  \begin{subfigure}[b]{0.49\linewidth}
    \centering
    \includegraphics[width=\linewidth]{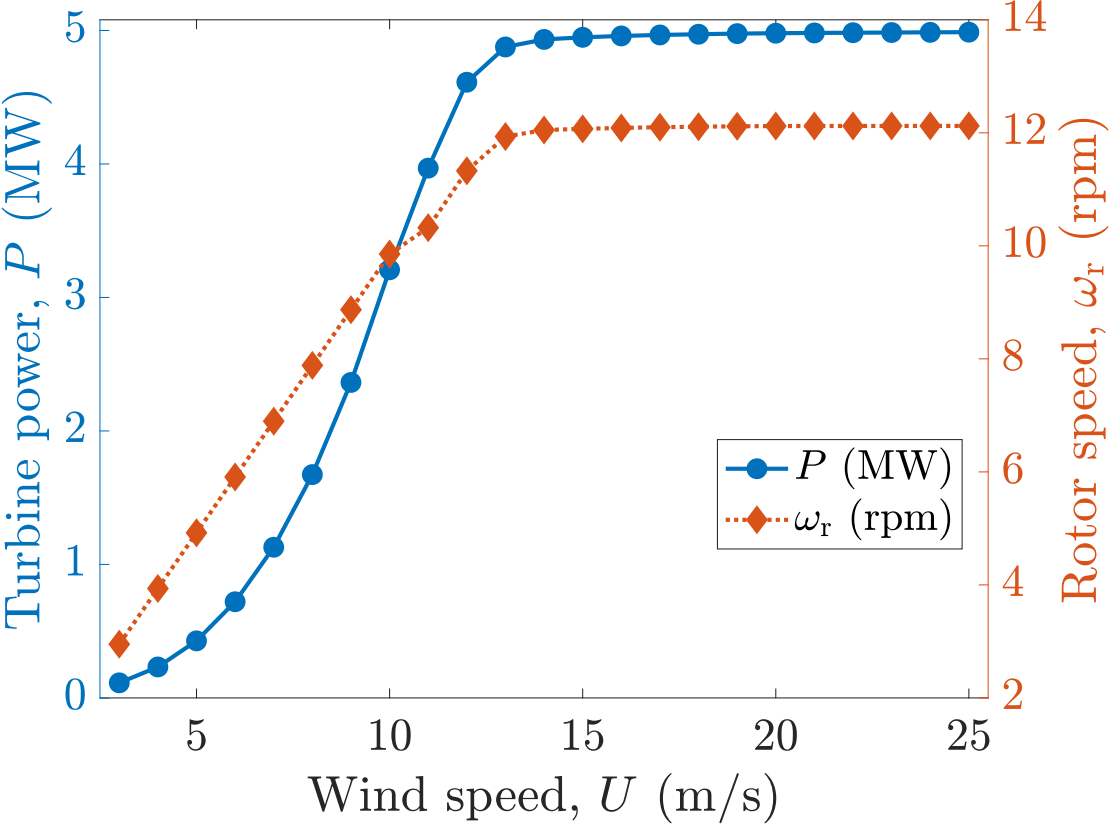}
    \caption{}
    \label{fig:sub2}
  \end{subfigure}
  %\caption{Performance curves of the wind turbine as functions of incident wind speed \(U\): mean electrical power output \(P\) (MW), blade pitch angle \(\theta_{\mathrm{blade}}\) (°), generator torque \(\tau_{g}\) (MN\(\cdot\)m), and rotor speed \(\omega_{r}\) (rpm). The computed average wind power is 1.92~MW, corresponding to an annual energy production (AEP) of 16.86~GWh.
%}
\caption{Performance curves of the wind turbine as functions of incident wind speed \(U\). 
(a) Blade pitch angle \(\theta_{\mathrm{blade}}\) (deg, left axis) and generator torque \(\tau_{g}\) (MN$\cdot$m, right axis). 
(b) Electrical power output \(P\) (MW, left axis) and rotor speed \(\omega_{r}\) (rpm, right axis). 
The computed average wind power is 1.92~MW, corresponding to an annual energy production (AEP) of 16.86~GWh.}

  \label{fig:wt_pwr_curve}
\end{figure}

In this section, we evaluate the Annual Energy Production (AEP) of both the wind turbine and the wave energy converters. To estimate the wind turbine AEP, we conducted a series of steady-state WEC-Sim simulations with constant wind speeds ranging from 5 to 25 m/s in 1 m/s increments. For each wind speed, we recorded the average electrical power output, blade pitch angle, generator torque, and rotor speed, as shown in Fig.~\ref{fig:wt_pwr_curve}. These simulations cover both Region 2 and Region 3 of wind turbine operation. As expected, the rated power of 5 MW is reached at approximately 12 m/s. In Region 3, the generator torque remains constant, while the rotor speed and power output exhibit only minor fluctuations. The blade pitch remains close to zero in Region 2 and increases with wind speed in Region 3. Based on these results, the estimated AEP of the wind turbine is 16.86 GWh, and the annual average power is 1.92 MW.

%\begin{figure}[ht!]
%    \centering
% \includegraphics[width=0.5\linewidth]{wt_pwr_curve.eps}   
%\end{figure}

\begin{figure}[ht!]
  \centering
  % (a)
  \begin{subfigure}[b]{0.49\linewidth}
    \centering
    \includegraphics[width=\linewidth]{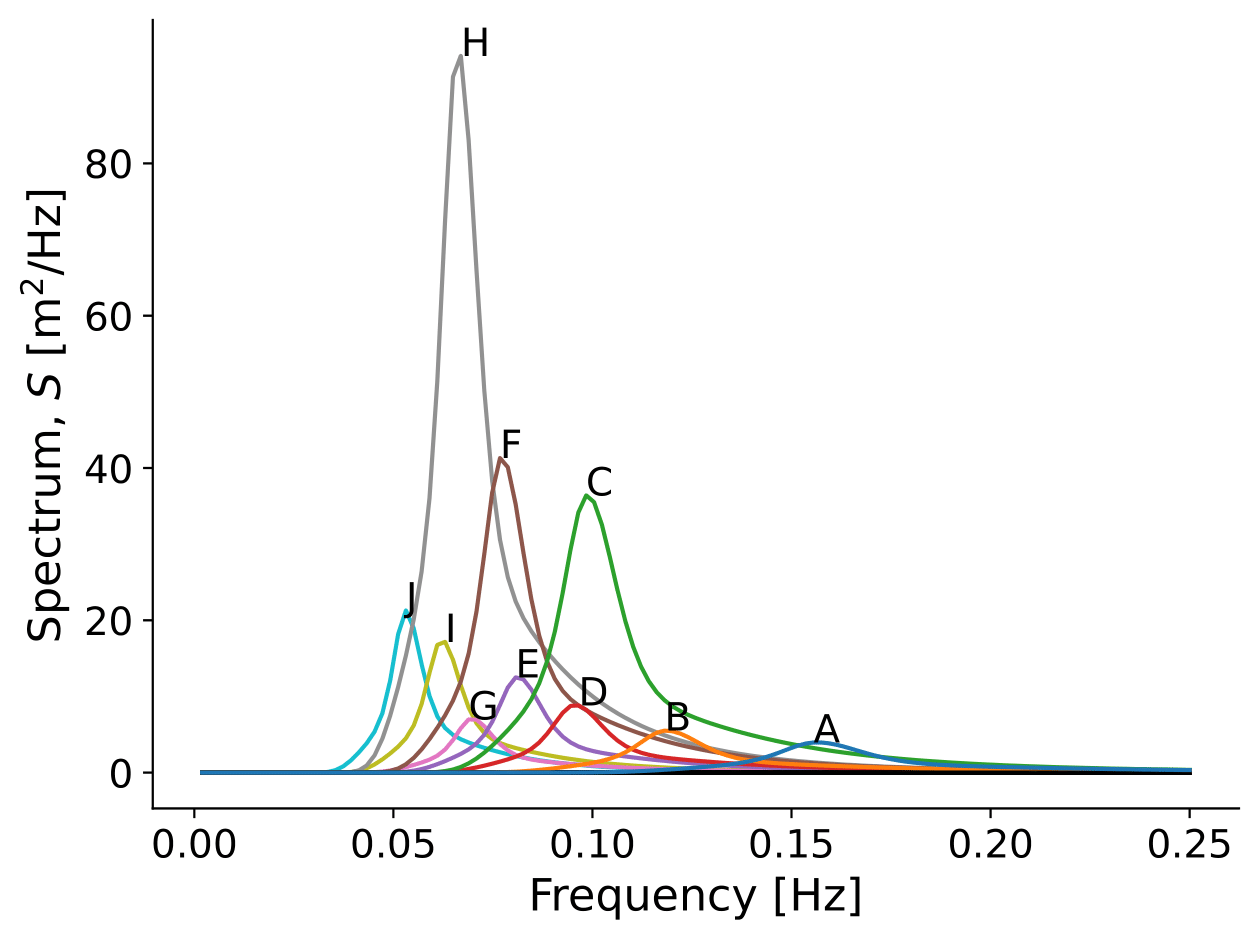}
    \caption{} \label{fig:wave_params}
  \end{subfigure}\hfill
  % (b)
  \begin{subfigure}[b]{0.49\linewidth}
    \centering
    \includegraphics[width=\linewidth]{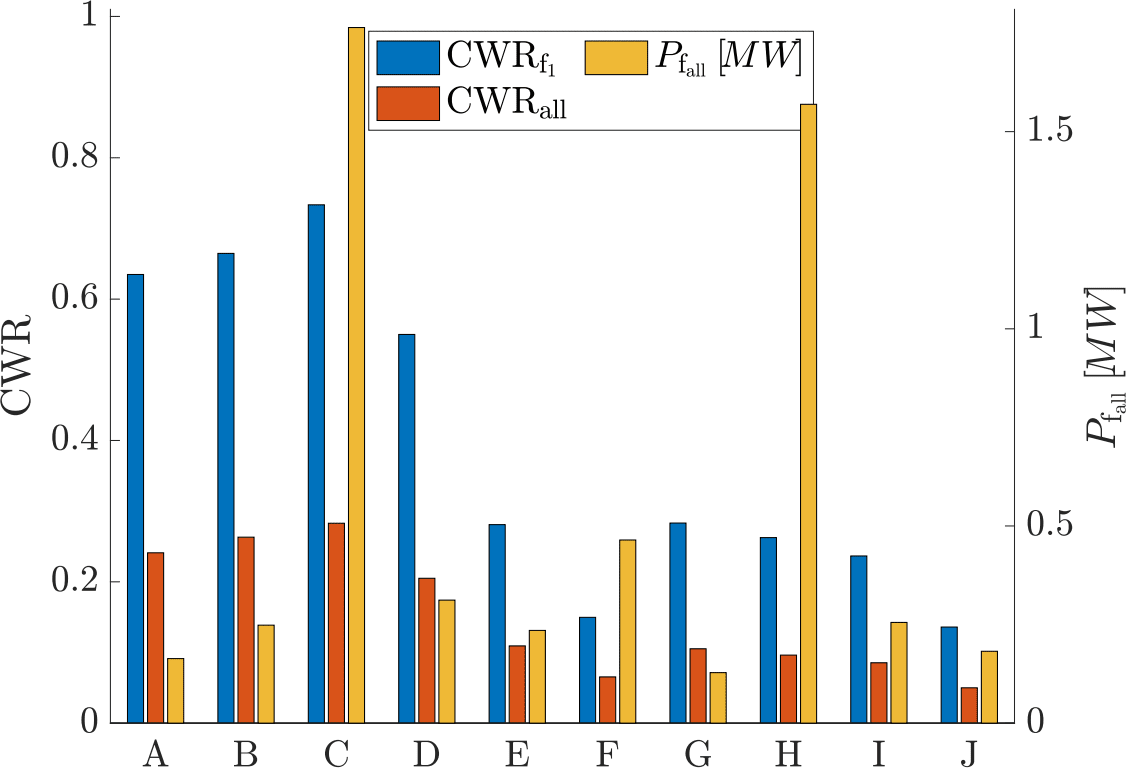}
    \caption{} \label{fig:pto_metrics}
  \end{subfigure}
\caption{(a) Significant wave height \(H_{m0}\) versus energy period \(T_e\) for the ten sea states (A–J) used in the WEC performance study.  
(b) Performance metrics across sea states A–J: capture width ratio of the first flap \(\mathrm{CWR}_{f1}\), total capture width ratio \(\mathrm{CWR}_{\mathrm{all}}\), and total extracted PTO power from all flaps \(P_{f_{\mathrm{all}}}\) [MW]. %, and mean PTO efficiency \(\eta_{f_{\mathrm{all}}}\). 
The annual weighted mean power across all sea states is 0.417~MW, which corresponds to an annual energy production (AEP) of 3.65~GWh.
}
  \label{fig:hybrid_results_wec}
\end{figure}

To evaluate the Annual Energy Production (AEP) of the wave energy converter (WEC), we used publicly available wave data from the CDIP-139 station located at the WEC site off the coast of Oregon, as previously shown in Fig.~\ref{fig:wave_data}(b).  Figure~\ref{fig:hybrid_results_wec}(a) displays the JONSWAP spectrum of the ten sea states considered in Fig.~\ref{fig:wave_data}(b), where the \(x\)-axis represents frequency (Hz) and the \(y\)-axis denotes the wave power spectrum \([{\mathrm{m}}^2/\mathrm{Hz}]\). For each of these ten sea states, we performed 10-minute WEC-Sim simulations, discarding the first 2 minutes to eliminate transient startup effects, and computed the average power extracted from the WEC. Figure~\ref{fig:hybrid_results_wec}(b) presents the simulation results. The first metric shown is the capture width ratio (CWR) of the first flap, which is calculated by dividing the average extracted power by the product of the wave power density and the flap width. This ratio quantifies how effectively the flap converts the incident wave energy into useful power. The second bar plot presents the total capture width ratio, calculated based on the combined power extracted by all three flaps. The third bar plot in Fig.~\ref{fig:hybrid_results_wec}(b) shows the total power extracted across the three flaps, computed as the sum of the time-averaged power from each flap. The total extracted power varies across sea states, ranging from approximately 0.12~MW to 1.7~MW.

%The third bar plot in Fig.~\ref{fig:hybrid_results_wec}(b) shows the mean power extracted across the three flaps, computed as the sum of the time-averaged power from each flap divided by three. The mean extracted power varies across sea states, ranging from approximately 0.06~MW to 0.59~MW. %The fourth bar plot displays the system efficiency, defined as the ratio of the total extracted power to the theoretical maximum. According to prior studies, the upper limit of the capture width ratio (CWR) for flap-type wave energy converters is approximately 50\% of the flap width~\cite{jin2022scalability}. Therefore, the efficiency is calculated by dividing the total extracted power by this theoretical maximum value.

%As observed, the efficiencies range from 5\% to 32\%, with noticeably lower values in sea states G through J. This decline can be attributed to a mismatch between the wave periods of these sea states and the flap’s natural period, which is approximately 4.2~s. In Fig.~\ref{fig:hybrid_results_wec}, the sea states are ordered by increasing wave period—from about 6~s in state A to nearly 17~s in state J—demonstrating that shorter-period waves generally yield higher efficiencies. This pattern suggests that greater deviation from the flap’s natural period reduces energy capture performance, especially in sea states E through J.

As seen in Fig.~\ref{fig:hybrid_results_wec}, the total extracted power peaks at sea states C and H. For sea state C, the high power output is primarily due to both high CWR and moderate wave power density, as previously illustrated in Fig.~\ref{fig:wave_data}(c). In contrast, for sea state H, although the CWR is relatively low, the wave power density is significantly higher, which compensates and results in substantial power extraction. 
Considering the possibility of occurrence of all sea states, the annual time-averaged power is approximately 0.417~MW, resulting in a WEC annual energy production (AEP) of 3.65~GWh. This corresponds to approximately 21.65\% of the wind turbine’s AEP (16.86~GWh), indicating that the WEC contributes about 17.8\% to the total annual energy production of the hybrid system. Notably, this additional energy contribution is achieved without a corresponding increase in submerged volume or material cost. As discussed earlier, the submerged volume of the hybrid platform is only about 57\% of that of the reference semi-submersible design. Therefore, the increased total energy output may be further enhanced while maintaining lower material usage and potentially reducing overall system cost.

\section{Limitations and Future Work}
\label{sec:lim_ftr_wrk}

This study establishes the foundational framework for the conceptual design, modeling, stability analysis, and performance assessment of a novel hybrid floating wind–wave energy platform. Nonetheless, several limitations persist, presenting valuable directions for future research.

\begin{itemize}
    \item \textbf{Structural Stress and Internal Load Modeling:}  
    The present study does not account for internal structural components required to satisfy stress constraints. Future work should incorporate reinforcements and subsystems, as these will influence system mass, distribution, and stability. External geometry may also require modification to meet material strength limits. Finite element analysis (FEA) tools such as Abaqus or ANSYS are recommended to model internal loads, identify stress concentrations, and assess structural integrity under coupled wind–wave conditions.

    \item \textbf{Coupled Hydrodynamics and Platform Dynamics:}  
    The current model captures linear hydrodynamic effects through potential flow theory but does not fully resolve nonlinear hydrodynamic interactions or coupled platform dynamics. Future work should incorporate detailed hydrodynamic coupling among the flaps and the central platform, as well as platform–structure coupling effects, to more accurately capture system behavior. Experimental validation and higher-fidelity simulations (e.g., CFD or nonlinear time-domain solvers) are also recommended to assess the influence of shadowing, viscous losses, and nonlinear wave loading on overall performance and stability.

    \item \textbf{Advanced Control Strategies:}  
The present study employs simplified control configurations for both the wind turbine and the wave energy converters (WECs). Future research should explore more advanced control strategies—such as open-loop optimal control, adaptive control, or model predictive control (MPC)—to enhance energy extraction and improve platform stability. Additionally, the control system should be designed using a holistic co-design framework that explicitly accounts for interactions between the wind and wave subsystems during the controller development process. This is particularly important, as the dynamic coupling between subsystems may significantly impact the effectiveness and robustness of the control strategy.

    \item \textbf{Lack of Optimization Framework:}  
    While the study includes a comprehensive sensitivity analysis to explore the influence of geometric parameters on performance metrics, no formal optimization was performed. The sensitivity results could serve as a basis for narrowing the design space and identifying promising regions for future optimization. Follow-up studies should implement control–co-design optimization frameworks to simultaneously refine geometric design, control laws, and performance objectives under site-specific environmental conditions.
\end{itemize}

\section{Conclusion}
\label{sec:cncl}
This study presents the conceptual design, modeling, and analysis of a novel hybrid floating offshore platform that integrates a wind turbine and three flap-type wave energy converters (WECs) within a unified hexagonal semi-submersible structure. Rather than treating the flaps as auxiliary devices added onto an existing wind platform, the hybrid concept was designed from the outset as a structurally and functionally integrated system, where the wind turbine, platform, and flap-type WECs are co-developed to share buoyancy, stability, and energy-harvesting functions. A consistent simulation and evaluation framework was developed to investigate hydrostatic stability, dynamic performance, and energy extraction under various environmental and design conditions. The design was benchmarked against the NREL 5~MW semisubmersible reference platform to ensure realistic scale and comparable dimensions. A metacentric height analysis confirmed that the system remains hydrostatically stable over a wide range of prescribed flap angles, particularly when using slurry ballast. A comprehensive sensitivity analysis revealed the influence of twelve geometric design variables on platform behavior and energy output. Flap dimensions and tower length were shown to strongly affect stability, WEC power, wind turbine performance, and structural stress.  

Time-domain simulations highlighted how wave incidence angle affects power sharing among the flaps, platform motions, and the resulting capture width ratio (CWR). In particular, the sensitivity study demonstrated that flaps aligned perpendicular to the incoming wave achieve higher efficiency, while shadowing effects reduce performance at oblique angles. Control strategies based on flap angle sweeps were shown to be effective in modulating pitch, offering an additional degree of freedom for stabilizing the system in steady-state conditions. Finally, annual energy production (AEP) calculations based on site-specific wind and wave conditions estimate 16.86~GWh for the wind turbine and 3.65~GWh for the WECs, with wave energy contributing approximately 17.8\% of the total hybrid energy output.  

Overall, the proposed hybrid system demonstrates the promise of an integrated offshore renewable energy platform that can ensure stability and deliver multi-functional energy harvesting. Future work should focus on structural stress modeling using finite element analysis, the development of advanced control strategies, and optimization-based co-design of geometry and control systems to fully realize the platform’s potential.

\bibliographystyle{asmejour}   %% .bst file that follows ASME journal format. Do not change.

\bibliography{Main} %% <=== change this to name of your bib file

%%%%%%%%%%%%%%%%%%%%%%%%%%%%%%%%%%%%%%%%%%%%%%%%%%%%%%%%%%%%%%%%%%%%%%

%% To omit final list of figures and tables, use the class option [nolists]

\end{document}